\documentclass[%
 reprint,
nofootinbib,
 amsmath,amssymb,
 aps,
]{revtex4-1}

\usepackage{graphicx}
\usepackage{dcolumn}
\usepackage{bm}
\usepackage{hyperref}

\usepackage{braket}
\usepackage{mathtools}
\usepackage{tikz}

\begin{document}

\preprint{APS/123-QED}

\title{Generation of n-qubit W states using Spin Torque}

\author{Amritesh Sharma}
 \email{amritesh.iitb@gmail.com}
\author{Ashwin A. Tulapurkar}%
 \email{ashwin@ee.iitb.ac.in}
\affiliation{%
Solid State Devices Group, Department of Electrical Engineering, Indian Institute of Technology, Bombay
}%

\date{\today}

\begin{abstract}
We examine here a scheme to generate a W state of an n-qubit system with all-to-all pairwise exchange interaction between n qubits. This relies on sharing of superposed excitations of a smaller number of $q$ qubits among others. We present a bound on the maximal jumps from q to n and formalize a scheme to generate $W_n$ state in $\mathcal{O}(\log_4 n)$ stages. We demonstrate this scheme in the context of spin torque based quantum computing architecture that are characterized by repeated interactions between static and flying qubits.


\end{abstract}

\pacs{Valid PACS appear here}
\maketitle


\section{\label{sec:intro} Introduction}

Quantum Entanglement is purely a non-classical phenomena that enables many Quantum Information Processing schemes and systems that exist today \citep{nielsen2010quantum, arute2019quantum, he2019two}.
Generation of any arbitrary entangled state is an important problem but certain entangled states are more useful. Two particularly inequivalent classes of tripartite entangled states viz Greenberger–Horne–Zeilinger (GHZ) and W exists \citep{dur2000three} and their generalizations to n-qubits have been extensively investigated \citep{teklemariam2002quantum, cabello2002bell, cabello2002two}.
W states particularly have garnered interest in the recent years because of their robustness against particle losses and local bit flips \citep{ccakmak2019robust, neven2018entanglement}.
$W_n$ states are defined as

\begin{equation} \label{eq:W_state}
 W_n = \frac{1}{\sqrt{n}}\sum_{i=1}^n |0_1 0_{2}... 1_i \,0_{i+1} ... 0_n>
\end{equation}

Applications exploiting these properties are not just limited to secure quantum communication \citep{liu2011efficient, zhu2015w, lipinska2018anonymous} and quantum teleportation \citep{zhao2004experimental, joo2003quantum}, ensemble based quantum memories have also been proposed using these states \citep{sangouard2011quantum}. They have also been used for addressing leader election problem \citep{d2004computational} , besides studying fundamentals of quantum mechanics \citep{teklemariam2002quantum}.
There have been proposals and some experimental demonstrations of generating these states in several competing Noisy Intermediate Scale (NISQ) \citep{Preskill2018quantumcomputingin} technologies of today like superconducting \citep{galiautdinov2012simple, kang2016fast, xiao2011generating, neeley2010generation, ccakmak2019robust}, photonic \citep{kiesel2003three, shi2002schemes, zou2002generation, li2004four, tashima2008elementary, heo2019scheme}, trapped ions \citep{haffner2005scalable, Sharma_2008}, 
etc.
Implementations based on a universal gate set of two qubit and single qubit operations can become complex as the system scales up \citep{yu2013optimal, diker2016deterministic}. Faster
and simpler methods would involve smaller number of architecture oriented gates and multi-qubit entanglement processes like shown in \citep{neeley2010generation}. Strategies exploiting lower order W states to create higher order W states can also be useful \citep{tashima2008elementary}. Besides, one circuit decomposition of an algorithm may not perform equally good on all implementations. So, architecture aware algorithms and hence circuit decomposition strategies is warranted
\citep{martinez2016compiling}.

Realization of qubits using localized spins is a promising technology \citep{bandyopadhyay2015introduction, he2019two}. Manipulation and control of spins is therefore very important.
'Classical' spin torque has played a key role in manipulating the magnetization of nano-magnets \citep{slonczewski1996current, berger1996emission}. When spin polarized current is injected into a ferromagnet (FM), the spin current polarized transverse to the magnetization direction of the FM is absorbed by it leading to a spin-torque. This is based on the exchange interaction between the conduction electrons and localized spins.  Various mechanism of producing spin polarized current such as spin-pumping \citep{bhuktare2019direct}, spin Hall effect \citep{bose2017sensitive, bose2018observation}, spin dependent thermoelectric effects \citep{bose2016observation}, spin Nernst effect \citep{bose2019recent, bose2018direct} etc. have been studied in detail. It has been shown that a quantum form of spin torque can be used for single and two qubit manipulations. This is based on the exchange interaction between static and flying qubits \citep{cordourier2010implementing}. There are many proposals for exploiting this scheme for applications in quantum information processing \citep{cordourier2010implementing, sutton2015manipulating, kulkarni2018transmission, kulkarni2019spin}.

We examine here a scheme to generate n-qubit W state in a system whose Hamiltonian takes a particular form: pair-wise exchange interaction between all the qubits. We show that time evolution with this Hamiltonian for a certain time followed by single qubit rotations lead to W state. We discuss the implementation of such a system in the context of a system of static and flying qubit where spin torque drives the evolution. The static qubits are assumed to be non-interacting, however repeated interactions with flying qubit can lead to effective exchange interaction between all the pairs of qubits.

\section{\label{sec:method} Method}

Consider a system of n 'spin-1/2's, each coupled with other via Heinsenberg exchange interaction so that the Hamiltonian can be written as:

\begin{equation} \label{eq:Ideal_Hamiltonain}
\mathcal{H} =J \sum_{i<j}  \bm{\sigma_i} \cdot \bm{\sigma_j}
\end{equation}

where $\bm{\sigma_i} = (\sigma_x, \sigma_y, \sigma_z)$ denote the respective Pauli operators. The basis states of each qubit denoted by $\ket{0}$ and $\ket{1}$, are eigenfunction of $\sigma_z$ with eigenvalues $\pm 1$ respectively. The Hamiltonian is block diagonal in the partitions of the computational basis ($\mathcal{B}^n$) where all states with fixed number of 0's and 1's are considered in one partition. Restricting to an ordered partition of one-hot-encoded states (i.e. states where only one spin is in state 1), denoted by $\mathcal{B}_1^n = \left\lbrace \ket{u_i} \in \mathcal{B}^n: \ket{u_i} = \ket{0_1...1_{n+1-i}...0_n} \right\rbrace $, the Hamiltonian can be written as

\begin{equation} \label{eq:HamiltonianAllSpinComb}
\mathcal{H} =J \left[ \frac{ (n-1)(n-4)}{2} \mathcal{I}_n+
\begin{bmatrix}
    0      & 2      & 2      & \dots \\
    2      & 0      & 2      & \dots \\
    \vdots & \vdots & \ddots & \vdots\\
    2      &      2 & \dots  & 0
\end{bmatrix} \right]
\end{equation}
where $\mathcal{I}_n$ denotes $n \times n$ identity matrix.

Note, $\mathcal{B}_1^n$ is closed under unitary evolution $U(t) = \exp(-i \mathcal{H} t)$ i.e. the state at time $t$, $\ket{\psi(t)}$ evolved from an initial state $\ket{\psi(0)}$ as $\ket{\psi(t)} = U(t) \ket{\psi(0)} \in \mathcal{B}_1^n $ if $\ket{\psi(0)} \in \mathcal{B}_1^n$. We take $\hbar =1$ in this paper. Now, if we start from an initial  unentangled state, say $ \ket{\psi(0)} = \ket{u_i}$, the state at time $t$ can be written as $ \ket{\psi(t)}  = a\ket{u_i} +b[ \ket{u_1} +...+ \ket{u_n} (\text{no} \ket{u_i} )  ]$.  The expressions for a and b are given as

\begin{equation}\label{eq:a_and_b}
\begin{split}
a &= \left[ \exp(i J nt) -i \frac{2}{n} \sin(J nt) \right] \exp(-i (n-2) J t) \\
b &= -i \frac{2}{n} \sin(J n t)\exp(-i (n-2) J t)
\end{split}
\end{equation}

The first term in Eq. \ref{eq:HamiltonianAllSpinComb}, which gives an overall phase to the time evolved wavefunction has been neglected in writing the above expressions. Turning off the Hamiltonian at an appropriate time $t = t_w$, we can reach an almost $W_n$ state (which we will call $\overline{W}_n$ from now) when $|b|^2=|a|^2=1/n$, with only exception of a relative phase factor given by $\exp(i \theta) = a/b$ between the initially excited and rest of the states in superposition. This can be corrected by applying a Z-axis rotation ($R_z$) through angle $\phi$ on the qubit which was excited in the initial state ($i^{th}$ qubit ) yielding the state $ a\exp(i \phi) \ket{u_i} +b[ \ket{u_1} +...+ \ket{u_n} (\text{no} \ket{u_i} ) ]$. Clearly, all coefficients would be in phase if $\phi = 2m\pi-\theta$ for an integer m thereby giving the desired $W_n$ state (Eq. \ref{eq:W_state})  up to a global phase. This idea of sharing a single excited qubit among the others has been previously adopted and shown experimentally for a system of 3 qubits \citep{neeley2010generation}. We run into problem extending this approach to larger number of qubits. This can be immediately noted by observing that $|b|^2 =1/n$ corresponds to condition, $\sin^2(Jnt_w) = n/4$ which cannot be satisfied for $n > 4$ for real $t$, that is to say this approach can be used to prepare $W_n$ state for $n=2,3 \; \text{and} \; 4$ only. The evolution time and z-axis rotation angle can be obtained from the expressions of a and b.

\begin{figure*}[tbp]
\includegraphics[width=0.95\textwidth, trim={0 0 0 0},clip]{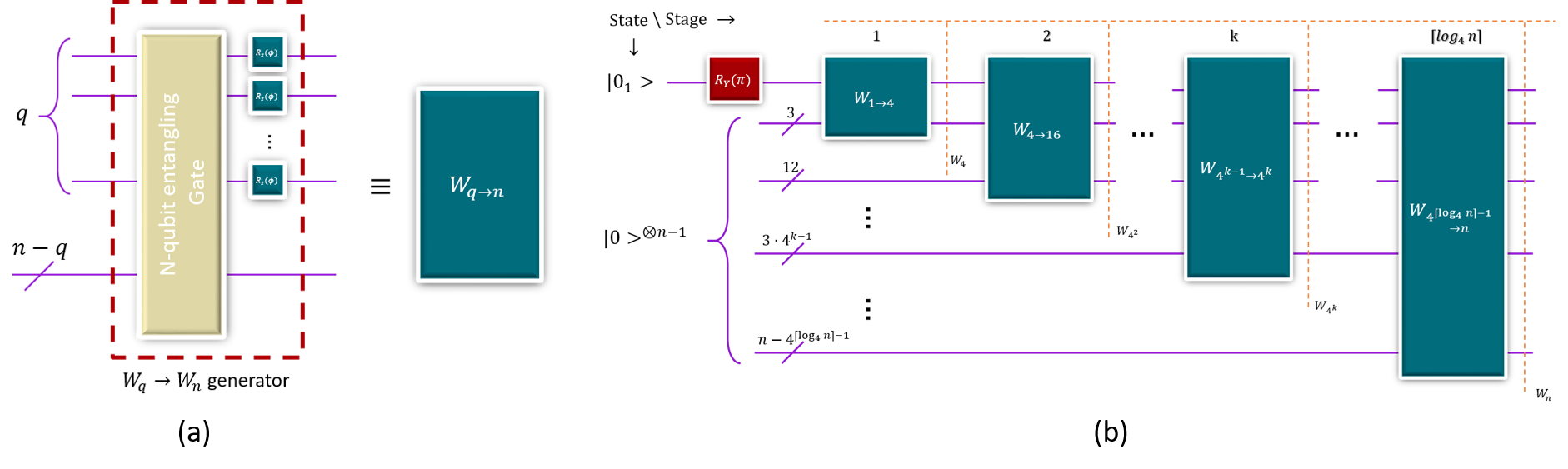}
\caption[]{ \textbf{$W_n$ state generation protocol}. \textbf{(a)} Wq to Wn generator block: n-qubit entangling operation followed by phase correction operations on q-qubits to generate a $W_n$ state from a $W_q$ state. \textbf{(b)} Circuit to generate $W_n$ state: Several Wq to Wn generator blocks are arranged successively in $\lceil \log_4 n \rceil$ to prepare a $W_n$ state starting from a $\ket{0}^{\otimes n}$ with one initial bit flip gate required to excite a single qubit.}
\label{fig:WnAlgorithm}
\end{figure*}


To overcome above limitation, we now show that staring from a $W_q$ state, it is possible to reach $W_{n }$ state where $n \leq 4q$. Consider an initial product state of q entangled qubits in state $W_q $ and $(n-q)$ qubits in state $\ket{0}$:

\begin{equation} \label{eq:ini_state}
\begin{split}
\ket{\psi(0)} & = \ket{W_q} \otimes \ket{0_{q+1} ... 0_{n}} \\
 & = \frac{1}{\sqrt{q}}\sum_{i=1}^{q} \ket{u_{n-i+1}}
\end{split}
\end{equation}

The time evolution of the above state is given by
$ \ket{\psi(t)}  = c\sum_{i=1}^{q} \ket{u_{n-i+1}} + d\sum_{i=q+1}^{n} \ket{u_{n-i+1}} $ where $c=\frac{a+(q-1)b}{\sqrt{q}}$
and $d=\frac{b q}{\sqrt{q}}=\sqrt{q} b $.
A necessary condition for it to be a W state is $|d|^2 =1/n$ which translates to $\sin^2\left(Jnt\right) = \frac{n}{4q}$. It has a real solution $t_w$ iff $n \leq 4q$. After reaching this state, the coupling is turned off, like before at time $t_w$, followed by correction of phase factors $\exp(i\theta) = c/d$ by single qubit operations $R_z(\phi)$ of either the first q qubits or last n-q qubits to reach the desired $W_{n}$ state.

The fact that from a $W_q$ state, we can jump only by 4x can also be seen by a different reasoning. The energy of the system i.e. $\langle H \rangle$ should be conserved during the time evolution. Neglecting the first (diagonal) term in Eq.~\ref{eq:HamiltonianAllSpinComb}, the matrix elements of Hamiltonian are given by, $ \bra{u_i} H \ket{u_j}=2J(1-\delta_{i,j})$. Using this, the energy of the state in Eq. \ref{eq:ini_state} is $E_0=2J(q-1)$. Let's assume that the state at time is $ \ket{\psi(t)}  = \frac{1}{\sqrt{n}}[exp(i\theta)\sum_{i=1}^{q} \ket{u_{n-i+1}} + \sum_{i=q+1}^{n} \ket{u_{n-i+1}}] $. (This corresponds to $\overline{W}_n$ state). The energy at time t is given by, $E=\frac{2J}{n} [2q(n-q)cos\phi +q(q-1)+(n-q)(n-q-1)]$. Equating $E$ to $E_0$ gives, $cos\theta=\frac{2q-n}{2q}$. This shows that $n$ can be atmost $4q$ and also gives an expression for the phase correction required to obtain $W_n$ state. The Hamiltonian in eqn. \ref{eq:HamiltonianAllSpinComb} commutes with total angular momentum, i.e. $[H,\bm{\sigma}]=0$, where $\bm{\sigma}=\sum_{i} \bm{\sigma_i}$. The basis states chosen are already eigenfunctions of $\bm{\sigma_z}$. The average value of $\bm{\sigma^2}$ should also be conserved during the time evolution. The Hamiltonian can be written as $H=(J/2)[\bm \sigma^2-\sum_i \bm \sigma_i^2]$. Thus conservation of $\bm{\sigma^2}$ is equivalent to conservation of energy. Therefore we can say that a jump more than 4x is not allowed as it does not conserve angular momentum.

This paves a way to generate an arbitrary $W_n$ state in $\mathcal{O}(\log_4 n)$ stages. In one possible path, every stage take a jump of 4x, starting from sharing a single excited state in stage 1 to reach a $W_{4^1}$ state, from which a $W_{4^2}$ state can be generated in stage 2 and so on to $W_{4^k}$ in kth stage assuming sufficient number of $\ket{0}$ polarized qubits are available in each stage. A maximum of $W_{4^{\lceil \log_4 n \rceil}}$ stage can be generated in the $\lceil \log_4 n \rceil$ stage which is sufficient for the required scheme since $n \leq 4^{\lceil \log_4 n \rceil}$ where $\lceil x \rceil$ indicates the smallest integer $\geq x$. One can alternatively think of generating W state backwards from a $W_{\lceil n/4 \rceil}$ state, which is generated from a $W_{\lceil n/4^2 \rceil}$ state and so on. Also, overhead cost of single qubit gates at kth stage for phase correction, which are operated in parallel, is $4^{k}$ which entails that total number of single qubit gates required would be $\mathcal{O}(4^{\lceil \log_4 n \rceil}) \approx \mathcal{O}(n)$. A schematic for the algorithm is  shown in Figure.\ref{fig:WnAlgorithm}.

\section{\label{sec:sTT} Spin Torque setting}

\begin{figure}[b]
\includegraphics[width=0.48\textwidth, trim={0 0 0 0},clip]{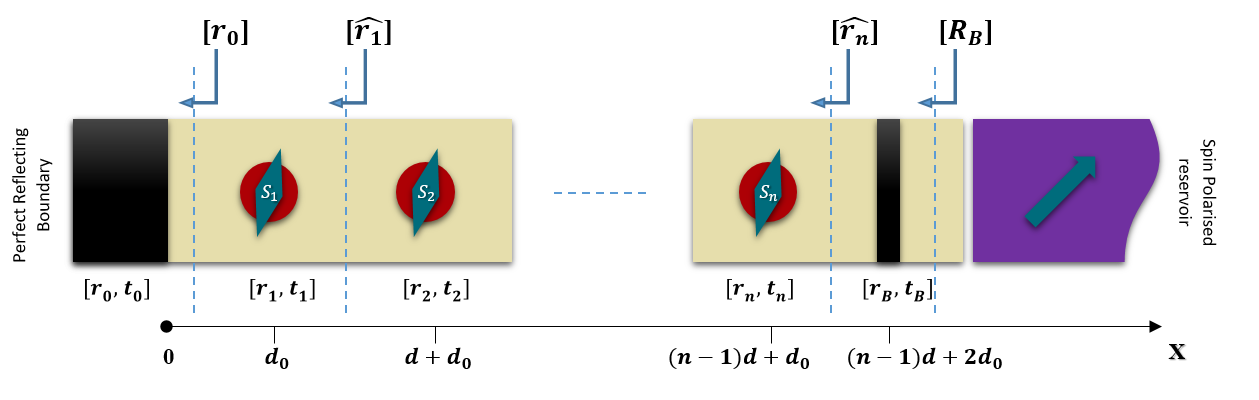}
\caption[]{\textbf{Schematic of the system}. n static qubits (colored red) in a spin coherent channel (shaded yellow). There are barrier gates (colored black) to facilitate creation of standing waves and a reservoir (colored purple) to inject and extract spin polarized carriers. The distance between two successive qubits is $d$ while that between a qubit and a barrier gate is $d_0$. Individual qubits act as spin-dependent scatterers with reflection and transmission denoted by $[r,t]$ matrices. Reflection matrices looking into the cascade of scatterers is also shown.}
\label{fig:QubitInChannel}
\end{figure}

In this work, we consider a system of impurity spins linearly embedded in a spin-coherent channel, placed sufficiently far so that there is no direct interaction between them. The separation between two consecutive impurity spins is taken to be all equal to d. As shown in the Figure.~\ref{fig:QubitInChannel}, on the left most side a hard barrier with perfect reflection is kept at a distance of $d_0$ from spin $S_1$. A barrier with partial reflection is present between the last spin $S_n$ and reservoir of electrons. The distance between $S_n$ and the barrier is taken as $d_0$. A reservoir held at desired spin potential is coupled to this channel that injects electrons into the channel. These itinerant spin carriers (flying qubits) collide with the immobile impurities (static qubits), get entangled with them and finally return back to the reservoir after multiple reflections thereby affecting a  spin-dependent rotation of system's state. This is equivalent to saying the rotation is caused by the application of quantum form of spin torque on the system of static qubits. Such successive collisions with impurities can be helpful to generate useful entanglement as we will show here.

The interaction between the flying and $j^{th}$ static qubit at location $x_j$, is governed by the exchange Hamiltonian

\begin{equation}
  \mathcal{H} = J_0 \bm{\sigma_f} \cdot \bm{\sigma_j} \delta(x-x_j)
\end{equation}

where $\bm{\sigma_f}$ and $\bm{\sigma_j}$ denote corresponding Pauli operators and $J_0$ the exchange strength. The transmission and reflection corresponding to this spin-dependent delta scatterer can be written as $t_j$ and $r_j = t_j - \mathcal{I}$ with

\begin{equation}
t_j = [\mathcal{I} + i \Omega \bm{\sigma_f} \cdot \bm{\sigma_j}]^{-1}
\end{equation}

where $\Omega = J_0/\hbar v$, $v$ is the velocity of injected electrons in the channel and $\mathcal{I}$ is identity matrix of dimension $2^{n+1} \times 2^{n+1}$. A cascade of $j^{th}$ scatterer from $(j-1)^{th}$ scatterer modifies its reflection as

\begin{equation} \label{eq:RefMatCascade}
\hat{r}_j = r_j + e^{2i kd_j} t_j \left(\mathcal{I} - e^{2i kd_j} \hat{r}_{j-1} r_j \right)^{-1} \hat{r}_{j-1} t_j
\end{equation}

where $k$ is the wave vector of injected electrons and $d_j$ is the distance of $j^{th}$ scatterer from $(j-1)^{th}$ scatterer.

These interactions are quite weak but can be enhanced with proper placement of static impurities 
and  barrier gates which facilitate formation of standing waves and thereby stronger interaction. These barrier gates have spin independent transmission
\begin{equation}
t_B = \frac{1}{(1 + i \Gamma)} \mathcal{I}
\end{equation}
and reflection given by $r_B = t_B - \mathcal{I}$. The hard barrier on the left most side is characterized by $r_B = - \mathcal{I}$ . The overall reflection matrix $\mathcal{R_B}$ for the cascade of such scatterers and barrier gates (c.f. Figure~\ref{fig:QubitInChannel}) can be obtained using Eq.~\ref{eq:RefMatCascade} as:

\begin{equation} \label{eq:final_RefMatCascade}
\mathcal{R_B} = r_B + e^{2i kd_0} t_B \left(\mathcal{I} - e^{2i kd_0} \hat{r}_{n} r_B \right)^{-1} \hat{r}_{n} t_B
\end{equation}

Such a method is described in \citep{sutton2015manipulating} for single and two qubit gates, and also discussed in \cite{kulkarni2018transmission, kulkarni2019spin} for various applications. We have here considered this method for more number of qubits between the two barrier gates and follow an iterative procedure for evolution \citep{sutton2015manipulating}. In this approach the role of time is taken by the number of flying qubits which have interacted with the static qubits. Denoting the state of the flying and n-qubit system at the beginning of $m^{th}$ iteration as $\rho_f$ and $\rho_s[m]$ we can write the initial separable state in the combined Hilbert space $ \mathcal{H}_f \otimes \mathcal{H}_1 \otimes \mathcal{H}_{2} \otimes \dots \mathcal{H}_n $ as $\rho_f \otimes \rho_s[m]$. The unitary matrix describing the overall reflection process $\mathcal{R_B}$, evolves it into $ \mathcal{R_B} \left( \rho_f \otimes \rho_s[m] \right) \mathcal{R_B}^\dagger $. The $(m+1)^{th}$ state of the n-qubit system can thus be obtained by taking a partial trace over the flying qubit's subspace.

\begin{equation} \label{eq:SystemEvolution}
\rho_s[m+1] = \text{Tr}_f \left[ \mathcal{R}_B (\rho_f \otimes \rho_s[m]) \mathcal{R}_B^\dagger \right]
\end{equation}

\subsection{\label{subsec:param} Choice of Parameters and Design Tradeoff}

We use reservoirs that inject $\ket {0}$ polarized electrons, in which case the relevant Kraus Operators for this evolution in the subspace of n-qubits, $\text{M}_k^{i,j} = \bra{k,i} {\mathcal{R}_B} \ket{0,j}$, are essentially the following partitions of the overall reflection matrix $\mathcal{R}_B$ in computational basis ${\mathcal{B}^{n+1}}$

\begin{subequations}
\begin{eqnarray}
\text{M}_0  = \mathcal{R_B}(1:2^n,1:2^n) \\
\text{M}_1  = \mathcal{R_B}(2^n+1:2^{n+1},1:2^n)
\end{eqnarray}
\end{subequations}

satisfying $\text{M}_0^\dagger \text{M}_0 + \text{M}_1^\dagger \text{M}_1 = \mathcal{I}_{2^n}$
such that the evolution of Eq.~\ref{eq:SystemEvolution} can be rephrased as

\begin{equation} \label{eq:SystemEvolutionKraus}
\rho_s[m+1] = \text{M}_0 \rho_s[m] \text{M}_0^\dagger + \text{M}_1 \rho_s[m] \text{M}_1^\dagger
\end{equation}

\begin{figure}[tbp]
\centering
\includegraphics[width=0.35\textwidth, trim={0 0 0 32},clip]{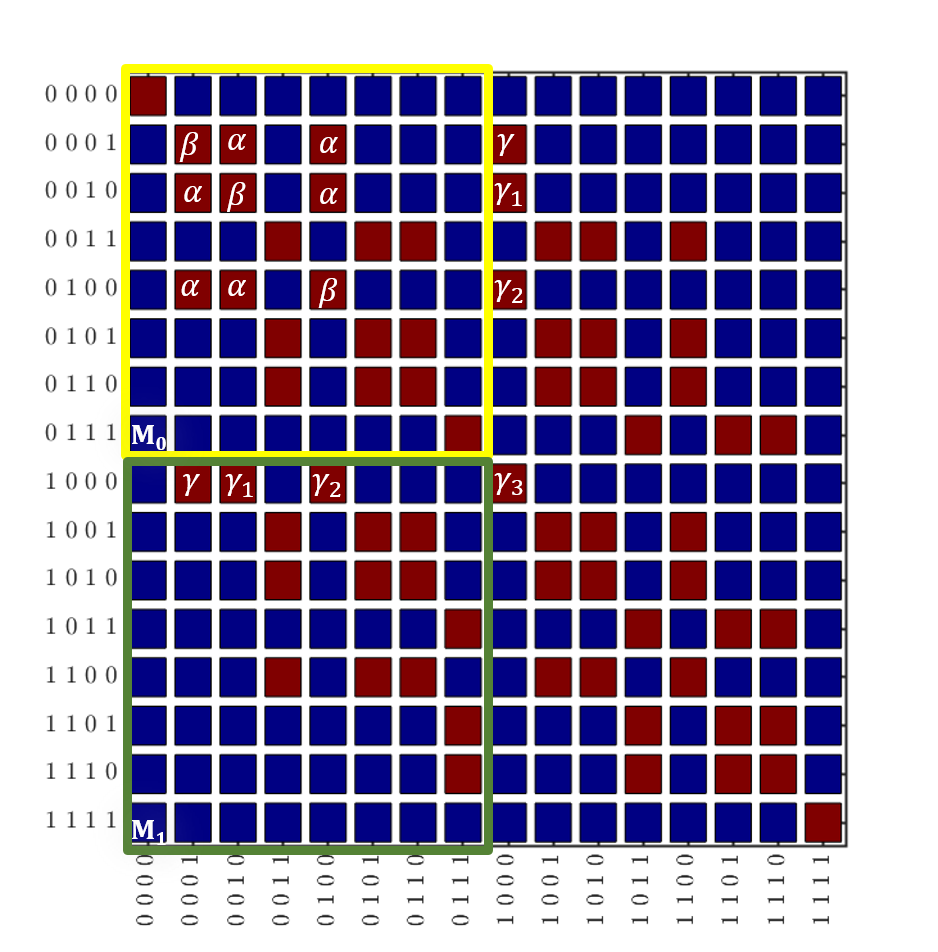}
\caption[] {\textbf{$\mathcal{R}_B$ for three qubits in the channel}. Non-zero elements of the overall Reflection Matrix $\mathcal{R}_B$ are colored red while the zero elements are colored blue independent of the parameters. The Kraus Operators $\text{M}_0$ and $\text{M}_1$ are partitions of the full $\mathcal{R}_B$ highlighted as yellow and green. The terms relevant to the evolution are $\alpha$ and $\beta$ as labelled while $\gamma$, $\gamma_1$, $\gamma_2$ and $\gamma_3$ give undesired superpositions in evolution.}
\label{fig:NonZeroReflectionMatrix}
\end{figure}

The elements of matrices $\text{M}_0$ and $\text{M}_1$ depend on the values of four parameters $kd$, $kd_0$, $\Gamma$ and $\Omega$ (since $d_j$ always appears together with $k$ we club them). We need to choose these values such that the evolution of static qubits given by Eq.~\ref{eq:SystemEvolutionKraus} can be approximated by unitary evolution corresponding to Hamiltonian in Eq.~\ref{eq:HamiltonianAllSpinComb}. For this, we analyze the system with three static qubit in channel. The optimized values of the parameters obtained are then used for all cases. Irrespective of the specific values of the aforementioned parameters, the operator $\mathcal{R}_B$ takes the form as shown in Figure.~\ref{fig:NonZeroReflectionMatrix}. We can identify the aforementioned Kraus Operators in the overall reflection matrix. Considering the one-hot encoded subspace of $M_0$ operator, we see that it can emulate the desired unitary, provided $ |\mathbf{\beta}|^2 + 2|\mathbf{\alpha}|^2 = 1$. The blocks of a block diagonal unitary operator are also unitary. Thus if we look for a parameter space where $\text{M}_0^\dagger \text{M}_0 \approx \mathcal{I}_{2^n}$ or $\text{M}_1^\dagger \text{M}_1 \approx \mathcal{O}_{2^n}$, the system would be evolved nearly unitarily by $\text{M}_0$ according to Eq.~\ref{eq:SystemEvolutionKraus}. This would enable to approximately realize the strategy described in previous section to generate $W_n$ state.

It is important to note that the off-diagonal term, '$\mathbf{\alpha}$', of $\text{M}_0$ in its one-hot encoded subspace is responsible for distributing the amplitude of an excited qubit among others. So, we should also try to maximize $|\mathbf{\alpha}|$ when looking for the optimal space besides trying to make $\text{M}_1^\dagger \text{M}_1 \approx \mathcal{O}_{2^n}$. We define a figure of merit (FOM), which we seek to minimize over all the parameters, as follows:

\begin{equation}
\text{FOM}_n = \log \left(\frac{ \left\lVert \text{M}_1 \right\rVert_F}{|\mathbf{\alpha}|} \right)
\end{equation}

where $ \left\lVert \text{M}_1 \right\rVert_F $ is the frobenius norm of the $\text{M}_1$ and n is number of qubits in channel.

\begin{figure}
\centering
\includegraphics[width=0.48\textwidth, trim={80 0 0 0},clip]{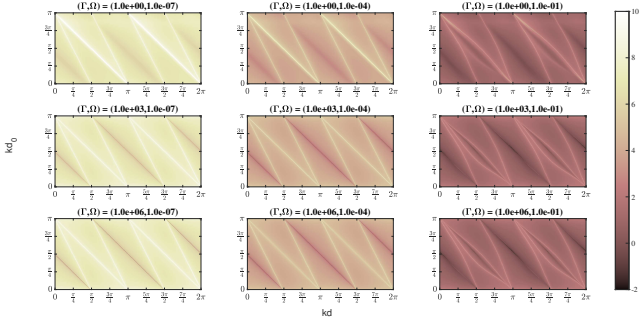}
\caption[] {\textbf{FOM$\bm{_3}$ as a function of $\bm{kd}$ and $\bm{kd_0}$} for different values of $(\Gamma,\Omega)$ about $ (1000, 0.0001)$. Good choices of $kd$ and $kd_0$ occur around a negative sloped line about $(\pi,\pi/2)$.
}
\label{fig:ReflectionMatrix_kd_vs_kd0}
\end{figure}

\begin{figure}
\centering
\includegraphics[width=0.48\textwidth, trim={80 0 0 0},clip]{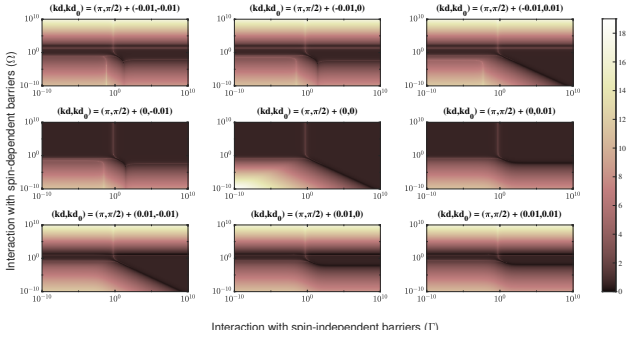}
\caption[] {\textbf{FOM$\bm{_3}$ as a function of $\bm{\Gamma}$ and $\bm{\Omega}$} for different choices of $(kd,kd_0)$ about $ (\pi, \pi/2)$.
}
\label{fig:ReflectionMatrix_Gamma_vs_Omega}
\end{figure}

Figure.~\ref{fig:ReflectionMatrix_kd_vs_kd0} shows the variation of  $\text{FOM}_3$ as a function of $kd$ and $kd_0$. A global minimum in this space occurs at $(kd,kd_0) = (\pi,\pi/2)$. The $\text{FOM}_3$ doesn't change much as long as we remain in negative slope region about the stated coordinate. Also, it gets affected only when we change $(\Gamma, \Omega)$ by orders of magnitude. So, it is quite robust against the choice of interaction strength parameters. Similarly Figure.~\ref{fig:ReflectionMatrix_Gamma_vs_Omega} shows the variation of  $\text{FOM}_3$ with $\Gamma$ and $\Omega$ corresponding to $1\%$ tolerance in the choices of $(kd,kd_0)$ about $(\pi,\pi/2)$. There are no common region of choices of $\Gamma$ and $\Omega$ among different plots implying $\text{FOM}_3$ is very sensitive to the choice of $(kd,kd_0)$. This  ascertains that given a geometry i.e. a choice of $(d,d_0)$ only electrons with certain wave vector can cause desired rotations which will be provided by the spin reservoirs connected to the channel. Our system implements single qubit rotations in the way discussed in \citep{sutton2015manipulating} for phase correction as discussed previously that requires either $\Omega$ or $kd_0$ to be small. This is contrary to the observation from Figure. \ref{fig:ReflectionMatrix_kd_vs_kd0} where a general decreasing trend for $\text{FOM}_3$ is observed with increasing $\Omega$. This implies a design tradeoff in the choice of parameters. Also, these parameters also affect the overall speed of the system in terms of total number of electrons required and further optimization can be done. We choose $(kd,kd_0) = (\pi,\pi/2)$ and $(\Gamma,\Omega) = (1000,0.0001)$ which are good enough parameters for the purpose of demonstration.

\begin{figure}
\centering
	\includegraphics[width=0.4\textwidth, trim={0 0 0 0},clip]{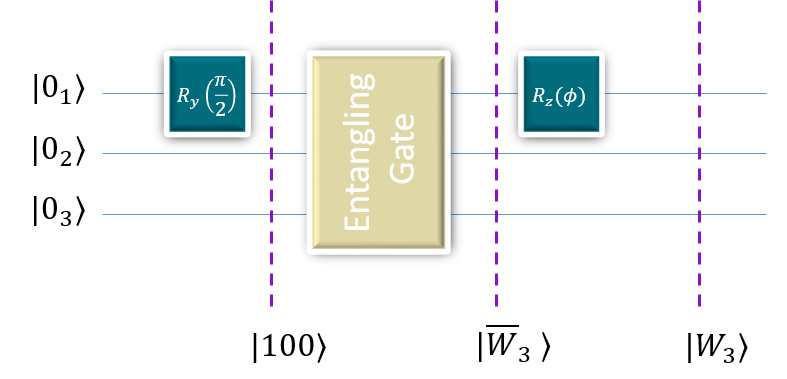}
\caption[] {\textbf{Quantum Circuit} for generation of $W_3$ state}
\label{fig:QuantumCircuitW3}
\end{figure}

\subsection{\label{subsec:evolution} Evolution}

\begin{figure}
\centering
	\includegraphics[width=0.45\textwidth, trim={0 0 0 0},clip]{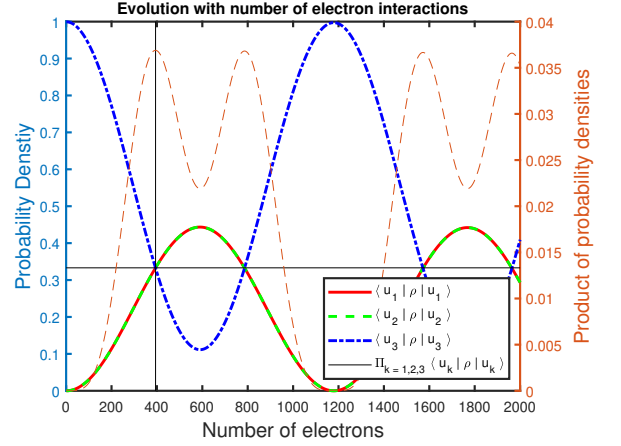}
\caption[] {\textbf{Three Qubit Evolution}. Evolution of Diagonal entries (left axis) and their product (right axis) of sytem's density matrix represented in $\mathcal{B}_1^3$. The vertical cursor marks the point when the product reaches the maximum suppossedly 1/27 in this case indicating the creation of $\overline{W}_3$ state. The horizontal cursor indicates the expected density matrix componenents value equal to 1/3.}
\label{fig:ThreeQubitEvolution}
\end{figure}

We will now illustrate the formation of $W_3$ state using the strategy highlighted in Section \ref{sec:method}. Consider the quantum circuit shown in Figure.~\ref{fig:QuantumCircuitW3}. Assume an initial state $\ket{000}$ for the three qubits in channel. Performing a rotation with y-polarized reservoir connected only to first qubit as prescribed in \citep{sutton2015manipulating}, we prepare a $\ket{100}$ state. After which we let open +z polarized reservoir injecting electrons in state $\ket{0}$, commonly connected to the spin chain system that enables an entangling evolution of the system governed by Eq.~\ref{eq:SystemEvolution}. Variation of diagonal components of the system's density matrix expressed in $\mathcal{B}_1^3$ with number of electrons while in this phase is shown in Figure.~\ref{fig:ThreeQubitEvolution}. We turn off the coupling to the reservoirs when the three curves intersect at a point to stop further evolution. The state of the system at this stage is actually $\overline{W}_3$. All diagonal components are nearly equal and ideally their product equals 1/27. As far as simulations are concerned we stop at the point when the product of diagonal entries reaches a maximum. Now, for the phase correction step we perform a single qubit manipulation of the first qubit like before but with z-polarized spins. This final evolution eventually leads to a $W_3$ state after which we close the gates.

To assess the quality of the $W_3$ state obtained we use fidelity that determines the closeness of two states. For a general state described by $\rho$ and a pure state $\ket{\psi}$ it is defined as $F(\rho, \ket{\psi}) = \sqrt{\bra{\psi} \rho \ket{\psi}}$.  In the scheme described above we stop the phase correction step when Fidelity between the evolving state and an ideal $W_n$ state reaches a maximum. For the three qubit case we obtain a Fidelity $\sim 99.9 \%$.

\begin{figure}
\centering
\includegraphics[width=0.48\textwidth, trim={0 0 0 0},clip]{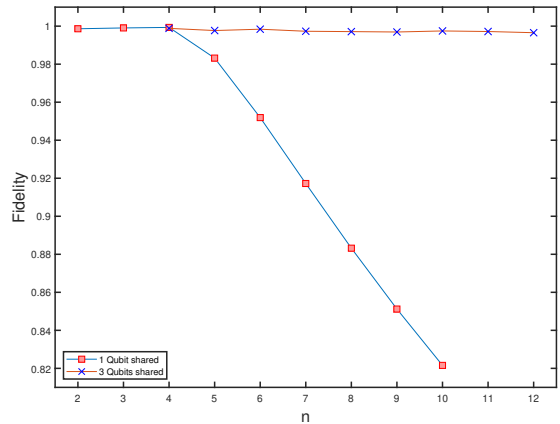}
\caption[] {\textbf{Fidelity of obtained states}. Fidelity of n-qubit W state obtained using single qubit sharing procedure vs sharing a superposition of three singly excited qubits (a $W_{q=3}$ state) as a function of n.}
\label{fig:FidelitySingleQubitExcitation}
\end{figure}

We follow the same procedure of sharing a singly excited qubit among others to generate a $W_n$ state with larger number of qubits in channel and observe that fidelity obtained decreases considerably beyond 10 qubits as shown in Figure.~\ref{fig:FidelitySingleQubitExcitation}. This is expected as this method of generating W state works only till $n=4$ as proved in section II. We obtain much better fidelities in preparing $W_n$ state using another $W_q$ state as the starting point. Figure.~\ref{fig:FidelitySingleQubitExcitation} also shows fidelities associated with creation of a $W_n$ state from $W_{q=3}$ state. We see that fidelities remain $\geq 99.9\%$ all the way upto $ n = 4q=12$ qubits.

\begin{figure}
\centering
\includegraphics[width=0.48\textwidth, trim={0 0 0 0},clip]{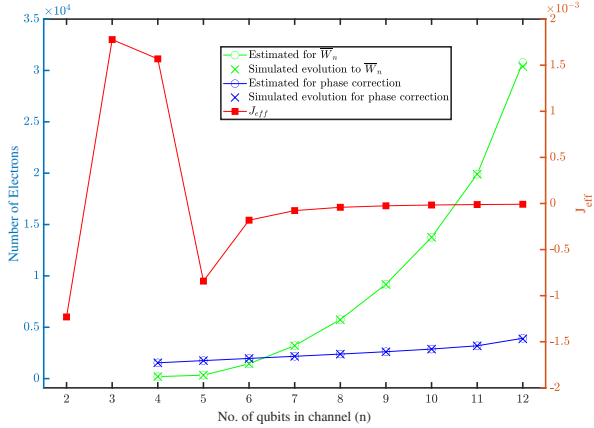}
\caption[] {\textbf{Number of electrons required for evolution}. Simulated number of electrons required towards entangling evolution for $\overline{W}_n$ state and for phase correction to get $W_n$ state starting from a $W_3$ state in the spin torque setting for $ 3 < n\leq 12$. Corresponding estimations shown as data circles match well with the simulations. Also, variation of $J_{\text{eff}}$ as a function of number of qubits shown against the right axis}
\label{fig:NumberOfElectronsRequired}
\end{figure}

In the present scheme, an electron undergoes multiple scattering at each spin site, as well as at the two barriers. As a result of this, we can engineer an effective interaction between all pairs of spins. The fact that we are able to generate W states with good fidelities indicate that the static qubit system, atleast in the one-hot-encoded space, is governed by the exchange Hamiltonian in Eq.~\ref{eq:HamiltonianAllSpinComb}. We can directly verify this by inspecting the $M_0$ matrix obtained numerically. If we assume that one electron interacts for a time $\delta t$ and causes a small rotation, then the evolution matrix is: $U(\delta t) = \exp(-i \mathcal{H} \delta t) = 1 - i \mathcal{H} \delta t$. We indeed checked that the relation $M_0 \approx 1 - i \mathcal{H} \delta t$ is satisfied in the one-hot-encoded subspace to a good accuracy. From this comparison we could also get the value of effective exchange constant $J_\text{eff}$ for unit $\delta t$, which is plotted in the Figure.~\ref{fig:NumberOfElectronsRequired} as a function of number of qubits. One can see oscillation in $J_\text{eff}$ followed by a decay as number of qubits increase. This is surprisingly reminiscent of the Ruderman–Kittel–Kasuya–Yosida (RKKY) oscillations in exchange coupling between two spins as a function of distance\citep{parkin1991spin}. The stopping time ($t_w$) and hence the number of electrons required (N) for obtaining $\overline{W}_{n}$ state starting from a $W_q$ is given by:
$\text{N}=\frac{t_w}{\delta t}=\frac{1}{ n J_{\text{eff}}(n)\delta t } \sin^{-1}\sqrt{\frac{n}{4q}}$. As shown in the Figure.~\ref{fig:NumberOfElectronsRequired}, the number of electrons obtained from this expression (green circle) and from complete numerical evolution (green cross) agree quite well. As the value of $J_\text{eff}$ decreases, the number of electrons required increase with the number of qubits. As discussed before, after reaching $\overline{W}_{n}$ state, we need single qubit rotations around z-axis to reach $W_n$ state. From the reflection matrix for single qubit operation, we can obtain the Kraus operator $\text{M}_0$ and compare it to the z-axis rotation operator $R_z(\phi)= \exp(-i \sigma_z \phi/2)$. This gives us the rotation per electron. The single qubit rotation required is determined from the phase factor $\exp({i \theta})=c/d$. From this we can estimate the number of electrons required for single qubit rotations. As shown in  Figure.~\ref{fig:NumberOfElectronsRequired}, this value (blue circle) agrees well with complete numerical evolution (blue cross).

Note that the number of electrons required increase as q to n jumps grow steeper (See green curve in Figure.~\ref{fig:NumberOfElectronsRequired}). The increase in number of electrons is due to the decrease in $J_\text{eff}$  for rotation of larger n. Thus, in this case non-maximal jumps may be beneficial in terms of number of electrons required. However if a system can be designed where $J_\text{eff}$ doesn't decrease as fast as $J_\text{eff}$ obtained for this choice of parameters, we could reach $W_n$ states quicker while taking maximum jumps.
It should be noted that during the transition from $W_q$ state to $\overline{W}_n$ state, the spin angular momentum of the static qubit system is conserved. During the transition from  $\overline{W}_n$ state to $W_n$ state the single qubit rotation operations change the spin angular momentum.

\begin{figure}
\centering
\includegraphics[width=0.48\textwidth, trim={30 10 20 0},clip]{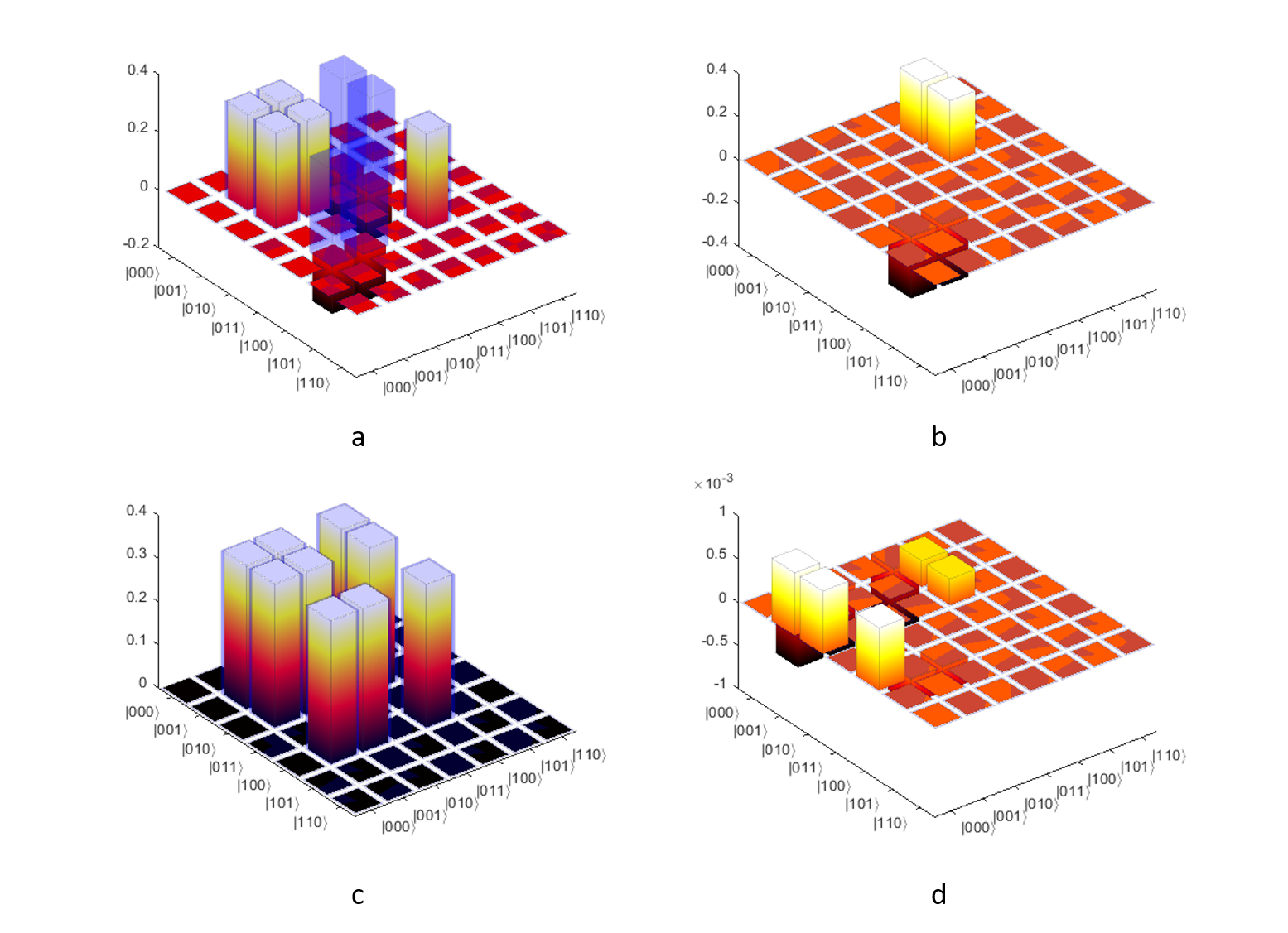}
\caption[] {\textbf{Density Matrix components}. Real (a,c) and imaginary (b,d) parts of obtained density matrix before and after phase correction respectively. The blue translucent and slightly thicker bars correspond to the density matrix of desired $W_3$ state. There is a good match between the obtained and expected desnsity matrices. The blue bars sheath the bars of real part of obtained denstiy matrix after phase correction with negligible error in imaginary part as also reflected by obtained Fidelity of $99.9\%$}
\label{fig:DMtomo}
\end{figure}

\section{\label{sec:arch} Modified Spin Torque Architecture}

Geometrically, there are two essential requirements on the architecture for this proposal. First, a random qubit must be uniquely accessible for performing single qubit manipulation. Second, it must allow n random qubits to be arranged in a one-dimensional fashion so that electrons from reservoirs can only interact successively for multi-qubit entanglement we have described above.

\begin{figure}[tbp]
\centering
\vspace{0.4cm}
\includegraphics[width=0.45\textwidth, trim={150 0 90 0},clip]{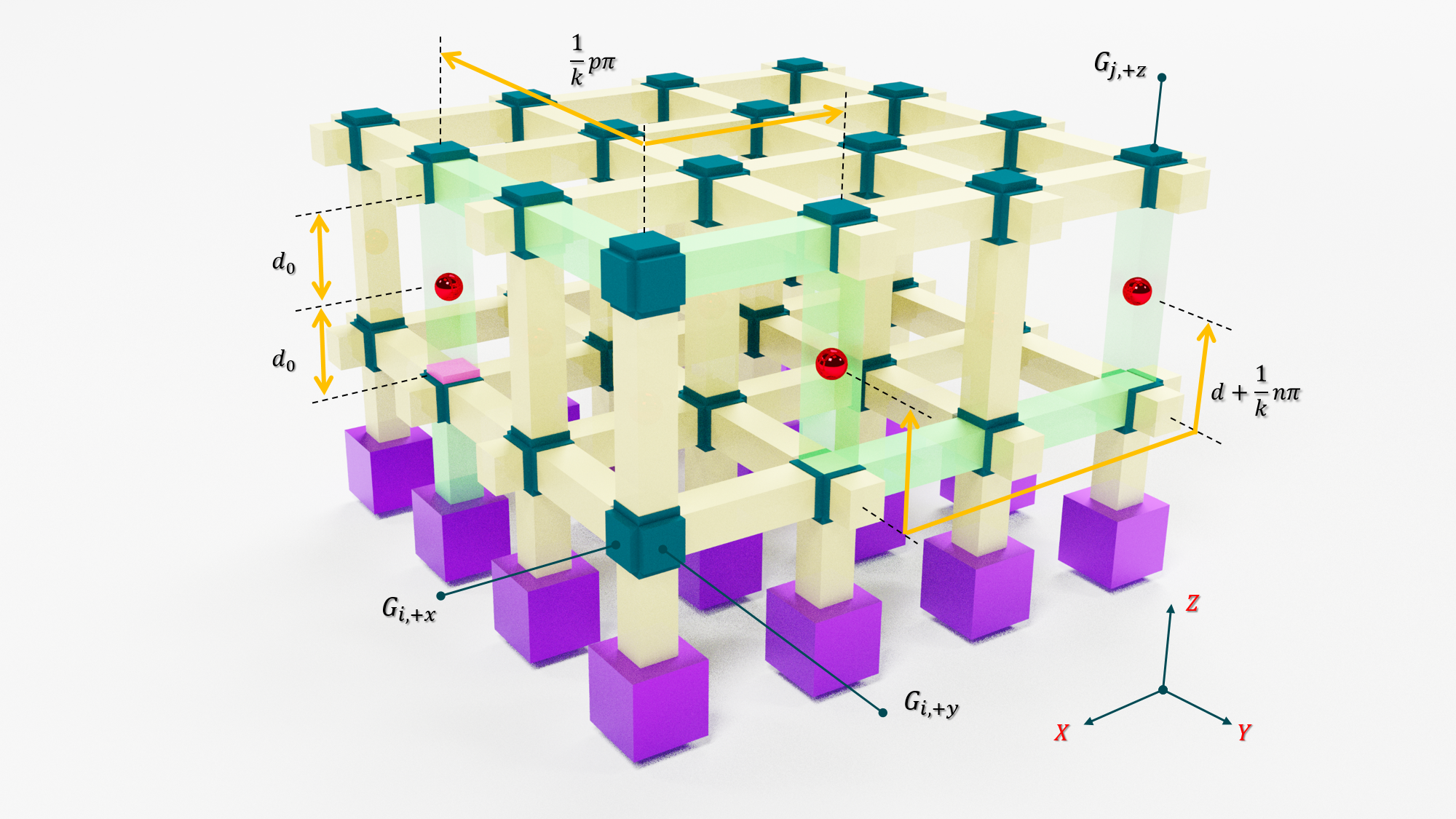}
\caption[] {\textbf{Schematic of the Proposed Architecture}. Static Qubits (red) embedded in the columns of spin-coherent channels (bright yellow) grid. Hard Barriers $G_{i,\hat{k}}$ (dark turquoise) guide the electrons along a path (transparent and green) shown for 3 qubit entanglement, partial barrier  for which are also shown (pink). Distance between a qubit and the gates facing it in each column is $d_0$ while the distance of the farther gates is roughly $d/2$. Manhattan distance between any two nodes on a horizontal level is integer multiple of $\pi/k$. This ensures any two qubits are separated effectively by a distance $d$ upto certain tolerances governed by optimization of parameters.}
\label{fig:Architecture}
\end{figure}

From Eq.~\ref{eq:RefMatCascade} we see that the evolution dynamics is unaffected if the choice of $d$ and $d_0$ are periodic with $\pi/k$. This enables us to stretch the one-dimensional arrangement of static qubits as shown in Figure.~\ref{fig:QubitInChannel} and create a more optimal arrangement of static qubits. Consider a meshed network of vertical and horizontal spin coherent semiconductor channels where a planar lattice of the static qubits is embedded in the columns of the network as shown in Figure.~\ref{fig:Architecture}. Here, we can uniquely access inidividual qubits through each column. To get n qubits in a line, we just need to 'needle' out a way through the channels as if the lattice were a two dimensional 'fabric'.

Each intersection 'i' of horizontal and vertical channels is provided with six barrier gates $G_{i,\hat{k}}$ where $\hat{k} = \pm x, \pm y, \pm z$ is the orientation of the face of the gate facing outward from the node. The potential of each gate is assumed to be fully controllable such that it can serve as a hard barrier, a partial one or a completely open. Hard barriers can be used for isolation of qubits when not operated upon and re-directing the flow of flying carriers between the horizontal and vertical channels besides when being used for perfect reflection as in the scheme proposed. Partial barriers (corresponds to $\Gamma$) are also required for facilitation of formation of standing waves to enhance the interaction of flying spins with static spins. The gates are required to be open while the flow of electrons is guided across an inter-sectional node. Flying electrons can be supplied from desired spin polarized reservoirs connected to the vertical columns.

There are a number of design considerations in this architecture similar to discussed in \citep{sutton2015manipulating}. We have assumed a $100\%$ polarization for the reservoirs which is seldom the case. However, it is not a primary requirement for the system to work, only that weaker polarization would incur more error in the evolution, as can be seen from Figure.~\ref{fig:NonZeroReflectionMatrix}, when undesired states outside the desired subspace states start sharing the superposition.
This arch is based on multiple reflections, so the value of $J_\text{eff}$ and its sign can be sensitive to the choice of k. Thus, proper reservoir engineering is necessary. Also, it requires long spin coherence length in the channels as the system scales up. Additionally, the gate control is much involved because of the proximity with other gates at each node, which partly overshadows the simplicity the scheme offers in the number of barrier gate switchings required as opposed to an all single and two-qubit universal gates based algorithm. This can become even more complex with single shot readout apparatus and requires further optimization in the design.

Note that this entire procedure can also be used to generate a pure superposition of all one-cold entangled states if we use down spin reservoirs where the evolution would happen in a one-cold encoded subspace. 
This can be equivalently seen by interchanging the roles of '0's and '1's. The architecture may also be utilized for preparation of generalized Dicke states and will be taken up in a future work.

\section{\label{sec:conclusion} Conclusion}
In summary we have presented a scheme to generate a $W_n$ state in systems with all-to-all exchange coupling between the constituent spins. We have shown that a single qubit sharing scheme works only upto four qubits and have suggested an improvement to start from another W state of smaller cardinality thereby outlining a procedure to generate a $W_n$ state in $\mathcal{O}(\log_4 n)$ stages. In the improved procedure only one-hot encoded subspace is utilized so physical systems which are equivalent in just this subspace also can be utilized in this scheme. We have shown that spin torque quantum computing architecture based on static and flying qubit interactions is one such avenue where it can be engineered i.e. the evolution operator can be made to emulate an all-to-all coupling unitary in a reduced subspace.

\begin{acknowledgments}
We acknowledge the support of Department of Science and Technology (DST), Government of India through Project No. SR/NM/NS-1112/2016 and Science and Engineering Research Board (SERB) through Project No. EMR/2016/007131.

\end{acknowledgments}


\bibliography{apssamp}

\providecommand{\noopsort}[1]{}\providecommand{\singleletter}[1]{#1}%
\begin{thebibliography}{46}%
\makeatletter
\providecommand \@ifxundefined [1]{%
 \@ifx{#1\undefined}
}%
\providecommand \@ifnum [1]{%
 \ifnum #1\expandafter \@firstoftwo
 \else \expandafter \@secondoftwo
 \fi
}%
\providecommand \@ifx [1]{%
 \ifx #1\expandafter \@firstoftwo
 \else \expandafter \@secondoftwo
 \fi
}%
\providecommand \natexlab [1]{#1}%
\providecommand \enquote  [1]{``#1''}%
\providecommand \bibnamefont  [1]{#1}%
\providecommand \bibfnamefont [1]{#1}%
\providecommand \citenamefont [1]{#1}%
\providecommand \href@noop [0]{\@secondoftwo}%
\providecommand \href [0]{\begingroup \@sanitize@url \@href}%
\providecommand \@href[1]{\@@startlink{#1}\@@href}%
\providecommand \@@href[1]{\endgroup#1\@@endlink}%
\providecommand \@sanitize@url [0]{\catcode `\\12\catcode `\$12\catcode
  `\&12\catcode `\#12\catcode `\^12\catcode `\_12\catcode `\%12\relax}%
\providecommand \@@startlink[1]{}%
\providecommand \@@endlink[0]{}%
\providecommand \url  [0]{\begingroup\@sanitize@url \@url }%
\providecommand \@url [1]{\endgroup\@href {#1}{\urlprefix }}%
\providecommand \urlprefix  [0]{URL }%
\providecommand \Eprint [0]{\href }%
\providecommand \doibase [0]{http://dx.doi.org/}%
\providecommand \selectlanguage [0]{\@gobble}%
\providecommand \bibinfo  [0]{\@secondoftwo}%
\providecommand \bibfield  [0]{\@secondoftwo}%
\providecommand \translation [1]{[#1]}%
\providecommand \BibitemOpen [0]{}%
\providecommand \bibitemStop [0]{}%
\providecommand \bibitemNoStop [0]{.\EOS\space}%
\providecommand \EOS [0]{\spacefactor3000\relax}%
\providecommand \BibitemShut  [1]{\csname bibitem#1\endcsname}%
\let\auto@bib@innerbib\@empty
\bibitem [{\citenamefont {Nielsen}\ and\ \citenamefont
  {Chuang}(2010)}]{nielsen2010quantum}%
  \BibitemOpen
  \bibfield  {author} {\bibinfo {author} {\bibfnamefont {M.~A.}\ \bibnamefont
  {Nielsen}}\ and\ \bibinfo {author} {\bibfnamefont {I.~L.}\ \bibnamefont
  {Chuang}},\ }\href@noop {} {\emph {\bibinfo {title} {Quantum Computation and
  Quantum Information}}}\ (\bibinfo  {publisher} {Cambridge University Press},\
  \bibinfo {year} {2010})\BibitemShut {NoStop}%
\bibitem [{\citenamefont {Arute}\ \emph {et~al.}(2019)\citenamefont {Arute},
  \citenamefont {Arya}, \citenamefont {Babbush}, \citenamefont {Bacon},
  \citenamefont {Bardin}, \citenamefont {Barends}, \citenamefont {Biswas},
  \citenamefont {Boixo}, \citenamefont {Brandao}, \citenamefont {Buell} \emph
  {et~al.}}]{arute2019quantum}%
  \BibitemOpen
  \bibfield  {author} {\bibinfo {author} {\bibfnamefont {F.}~\bibnamefont
  {Arute}}, \bibinfo {author} {\bibfnamefont {K.}~\bibnamefont {Arya}},
  \bibinfo {author} {\bibfnamefont {R.}~\bibnamefont {Babbush}}, \bibinfo
  {author} {\bibfnamefont {D.}~\bibnamefont {Bacon}}, \bibinfo {author}
  {\bibfnamefont {J.~C.}\ \bibnamefont {Bardin}}, \bibinfo {author}
  {\bibfnamefont {R.}~\bibnamefont {Barends}}, \bibinfo {author} {\bibfnamefont
  {R.}~\bibnamefont {Biswas}}, \bibinfo {author} {\bibfnamefont
  {S.}~\bibnamefont {Boixo}}, \bibinfo {author} {\bibfnamefont {F.~G.}\
  \bibnamefont {Brandao}}, \bibinfo {author} {\bibfnamefont {D.~A.}\
  \bibnamefont {Buell}},  \emph {et~al.},\ }\href@noop {} {\bibfield  {journal}
  {\bibinfo  {journal} {Nature}\ }\textbf {\bibinfo {volume} {574}},\ \bibinfo
  {pages} {505} (\bibinfo {year} {2019})}\BibitemShut {NoStop}%
\bibitem [{\citenamefont {He}\ \emph {et~al.}(2019)\citenamefont {He},
  \citenamefont {Gorman}, \citenamefont {Keith}, \citenamefont {Kranz},
  \citenamefont {Keizer},\ and\ \citenamefont {Simmons}}]{he2019two}%
  \BibitemOpen
  \bibfield  {author} {\bibinfo {author} {\bibfnamefont {Y.}~\bibnamefont
  {He}}, \bibinfo {author} {\bibfnamefont {S.}~\bibnamefont {Gorman}}, \bibinfo
  {author} {\bibfnamefont {D.}~\bibnamefont {Keith}}, \bibinfo {author}
  {\bibfnamefont {L.}~\bibnamefont {Kranz}}, \bibinfo {author} {\bibfnamefont
  {J.}~\bibnamefont {Keizer}}, \ and\ \bibinfo {author} {\bibfnamefont
  {M.}~\bibnamefont {Simmons}},\ }\href@noop {} {\bibfield  {journal} {\bibinfo
   {journal} {Nature}\ }\textbf {\bibinfo {volume} {571}},\ \bibinfo {pages}
  {371} (\bibinfo {year} {2019})}\BibitemShut {NoStop}%
\bibitem [{\citenamefont {D{\"u}r}\ \emph {et~al.}(2000)\citenamefont
  {D{\"u}r}, \citenamefont {Vidal},\ and\ \citenamefont
  {Cirac}}]{dur2000three}%
  \BibitemOpen
  \bibfield  {author} {\bibinfo {author} {\bibfnamefont {W.}~\bibnamefont
  {D{\"u}r}}, \bibinfo {author} {\bibfnamefont {G.}~\bibnamefont {Vidal}}, \
  and\ \bibinfo {author} {\bibfnamefont {J.~I.}\ \bibnamefont {Cirac}},\
  }\href@noop {} {\bibfield  {journal} {\bibinfo  {journal} {Physical Review
  A}\ }\textbf {\bibinfo {volume} {62}},\ \bibinfo {pages} {062314} (\bibinfo
  {year} {2000})}\BibitemShut {NoStop}%
\bibitem [{\citenamefont {Teklemariam}\ \emph {et~al.}(2002)\citenamefont
  {Teklemariam}, \citenamefont {Fortunato}, \citenamefont {Pravia},
  \citenamefont {Sharf}, \citenamefont {Havel}, \citenamefont {Cory},
  \citenamefont {Bhattaharyya},\ and\ \citenamefont
  {Hou}}]{teklemariam2002quantum}%
  \BibitemOpen
  \bibfield  {author} {\bibinfo {author} {\bibfnamefont {G.}~\bibnamefont
  {Teklemariam}}, \bibinfo {author} {\bibfnamefont {E.}~\bibnamefont
  {Fortunato}}, \bibinfo {author} {\bibfnamefont {M.}~\bibnamefont {Pravia}},
  \bibinfo {author} {\bibfnamefont {Y.}~\bibnamefont {Sharf}}, \bibinfo
  {author} {\bibfnamefont {T.}~\bibnamefont {Havel}}, \bibinfo {author}
  {\bibfnamefont {D.}~\bibnamefont {Cory}}, \bibinfo {author} {\bibfnamefont
  {A.}~\bibnamefont {Bhattaharyya}}, \ and\ \bibinfo {author} {\bibfnamefont
  {J.}~\bibnamefont {Hou}},\ }\href@noop {} {\bibfield  {journal} {\bibinfo
  {journal} {Physical Review A}\ }\textbf {\bibinfo {volume} {66}},\ \bibinfo
  {pages} {012309} (\bibinfo {year} {2002})}\BibitemShut {NoStop}%
\bibitem [{\citenamefont {Cabello}(2002{\natexlab{a}})}]{cabello2002bell}%
  \BibitemOpen
  \bibfield  {author} {\bibinfo {author} {\bibfnamefont {A.}~\bibnamefont
  {Cabello}},\ }\href@noop {} {\bibfield  {journal} {\bibinfo  {journal}
  {Physical Review A}\ }\textbf {\bibinfo {volume} {65}},\ \bibinfo {pages}
  {032108} (\bibinfo {year} {2002}{\natexlab{a}})}\BibitemShut {NoStop}%
\bibitem [{\citenamefont {Cabello}(2002{\natexlab{b}})}]{cabello2002two}%
  \BibitemOpen
  \bibfield  {author} {\bibinfo {author} {\bibfnamefont {A.}~\bibnamefont
  {Cabello}},\ }\href@noop {} {\bibfield  {journal} {\bibinfo  {journal}
  {Physical Review A}\ }\textbf {\bibinfo {volume} {66}},\ \bibinfo {pages}
  {042114} (\bibinfo {year} {2002}{\natexlab{b}})}\BibitemShut {NoStop}%
\bibitem [{\citenamefont {{\c{C}}akmak}\ \emph {et~al.}(2019)\citenamefont
  {{\c{C}}akmak}, \citenamefont {Campbell}, \citenamefont {Vacchini},
  \citenamefont {M{\"u}stecapl{\i}o{\u{g}}lu},\ and\ \citenamefont
  {Paternostro}}]{ccakmak2019robust}%
  \BibitemOpen
  \bibfield  {author} {\bibinfo {author} {\bibfnamefont {B.}~\bibnamefont
  {{\c{C}}akmak}}, \bibinfo {author} {\bibfnamefont {S.}~\bibnamefont
  {Campbell}}, \bibinfo {author} {\bibfnamefont {B.}~\bibnamefont {Vacchini}},
  \bibinfo {author} {\bibfnamefont {{\"O}.~E.}\ \bibnamefont
  {M{\"u}stecapl{\i}o{\u{g}}lu}}, \ and\ \bibinfo {author} {\bibfnamefont
  {M.}~\bibnamefont {Paternostro}},\ }\href@noop {} {\bibfield  {journal}
  {\bibinfo  {journal} {Physical Review A}\ }\textbf {\bibinfo {volume} {99}},\
  \bibinfo {pages} {012319} (\bibinfo {year} {2019})}\BibitemShut {NoStop}%
\bibitem [{\citenamefont {Neven}\ \emph {et~al.}(2018)\citenamefont {Neven},
  \citenamefont {Martin},\ and\ \citenamefont
  {Bastin}}]{neven2018entanglement}%
  \BibitemOpen
  \bibfield  {author} {\bibinfo {author} {\bibfnamefont {A.}~\bibnamefont
  {Neven}}, \bibinfo {author} {\bibfnamefont {J.}~\bibnamefont {Martin}}, \
  and\ \bibinfo {author} {\bibfnamefont {T.}~\bibnamefont {Bastin}},\
  }\href@noop {} {\bibfield  {journal} {\bibinfo  {journal} {Physical Review
  A}\ }\textbf {\bibinfo {volume} {98}},\ \bibinfo {pages} {062335} (\bibinfo
  {year} {2018})}\BibitemShut {NoStop}%
\bibitem [{\citenamefont {Liu}\ \emph {et~al.}(2011)\citenamefont {Liu},
  \citenamefont {Wang},\ and\ \citenamefont {Jiang}}]{liu2011efficient}%
  \BibitemOpen
  \bibfield  {author} {\bibinfo {author} {\bibfnamefont {W.}~\bibnamefont
  {Liu}}, \bibinfo {author} {\bibfnamefont {Y.-B.}\ \bibnamefont {Wang}}, \
  and\ \bibinfo {author} {\bibfnamefont {Z.-T.}\ \bibnamefont {Jiang}},\
  }\href@noop {} {\bibfield  {journal} {\bibinfo  {journal} {Optics
  Communications}\ }\textbf {\bibinfo {volume} {284}},\ \bibinfo {pages} {3160}
  (\bibinfo {year} {2011})}\BibitemShut {NoStop}%
\bibitem [{\citenamefont {Zhu}\ \emph {et~al.}(2015)\citenamefont {Zhu},
  \citenamefont {Xu},\ and\ \citenamefont {Pei}}]{zhu2015w}%
  \BibitemOpen
  \bibfield  {author} {\bibinfo {author} {\bibfnamefont {C.}~\bibnamefont
  {Zhu}}, \bibinfo {author} {\bibfnamefont {F.}~\bibnamefont {Xu}}, \ and\
  \bibinfo {author} {\bibfnamefont {C.}~\bibnamefont {Pei}},\ }\href@noop {}
  {\bibfield  {journal} {\bibinfo  {journal} {Scientific reports}\ }\textbf
  {\bibinfo {volume} {5}},\ \bibinfo {pages} {17449} (\bibinfo {year}
  {2015})}\BibitemShut {NoStop}%
\bibitem [{\citenamefont {Lipinska}\ \emph {et~al.}(2018)\citenamefont
  {Lipinska}, \citenamefont {Murta},\ and\ \citenamefont
  {Wehner}}]{lipinska2018anonymous}%
  \BibitemOpen
  \bibfield  {author} {\bibinfo {author} {\bibfnamefont {V.}~\bibnamefont
  {Lipinska}}, \bibinfo {author} {\bibfnamefont {G.}~\bibnamefont {Murta}}, \
  and\ \bibinfo {author} {\bibfnamefont {S.}~\bibnamefont {Wehner}},\
  }\href@noop {} {\bibfield  {journal} {\bibinfo  {journal} {Physical Review
  A}\ }\textbf {\bibinfo {volume} {98}},\ \bibinfo {pages} {052320} (\bibinfo
  {year} {2018})}\BibitemShut {NoStop}%
\bibitem [{\citenamefont {Zhao}\ \emph {et~al.}(2004)\citenamefont {Zhao},
  \citenamefont {Chen}, \citenamefont {Zhang}, \citenamefont {Yang},
  \citenamefont {Briegel},\ and\ \citenamefont {Pan}}]{zhao2004experimental}%
  \BibitemOpen
  \bibfield  {author} {\bibinfo {author} {\bibfnamefont {Z.}~\bibnamefont
  {Zhao}}, \bibinfo {author} {\bibfnamefont {Y.-A.}\ \bibnamefont {Chen}},
  \bibinfo {author} {\bibfnamefont {A.-N.}\ \bibnamefont {Zhang}}, \bibinfo
  {author} {\bibfnamefont {T.}~\bibnamefont {Yang}}, \bibinfo {author}
  {\bibfnamefont {H.~J.}\ \bibnamefont {Briegel}}, \ and\ \bibinfo {author}
  {\bibfnamefont {J.-W.}\ \bibnamefont {Pan}},\ }\href@noop {} {\bibfield
  {journal} {\bibinfo  {journal} {Nature}\ }\textbf {\bibinfo {volume} {430}},\
  \bibinfo {pages} {54} (\bibinfo {year} {2004})}\BibitemShut {NoStop}%
\bibitem [{\citenamefont {Joo}\ \emph {et~al.}(2003)\citenamefont {Joo},
  \citenamefont {Park}, \citenamefont {Oh},\ and\ \citenamefont
  {Kim}}]{joo2003quantum}%
  \BibitemOpen
  \bibfield  {author} {\bibinfo {author} {\bibfnamefont {J.}~\bibnamefont
  {Joo}}, \bibinfo {author} {\bibfnamefont {Y.-J.}\ \bibnamefont {Park}},
  \bibinfo {author} {\bibfnamefont {S.}~\bibnamefont {Oh}}, \ and\ \bibinfo
  {author} {\bibfnamefont {J.}~\bibnamefont {Kim}},\ }\href@noop {} {\bibfield
  {journal} {\bibinfo  {journal} {New Journal of Physics}\ }\textbf {\bibinfo
  {volume} {5}},\ \bibinfo {pages} {136} (\bibinfo {year} {2003})}\BibitemShut
  {NoStop}%
\bibitem [{\citenamefont {Sangouard}\ \emph {et~al.}(2011)\citenamefont
  {Sangouard}, \citenamefont {Simon}, \citenamefont {De~Riedmatten},\ and\
  \citenamefont {Gisin}}]{sangouard2011quantum}%
  \BibitemOpen
  \bibfield  {author} {\bibinfo {author} {\bibfnamefont {N.}~\bibnamefont
  {Sangouard}}, \bibinfo {author} {\bibfnamefont {C.}~\bibnamefont {Simon}},
  \bibinfo {author} {\bibfnamefont {H.}~\bibnamefont {De~Riedmatten}}, \ and\
  \bibinfo {author} {\bibfnamefont {N.}~\bibnamefont {Gisin}},\ }\href@noop {}
  {\bibfield  {journal} {\bibinfo  {journal} {Reviews of Modern Physics}\
  }\textbf {\bibinfo {volume} {83}},\ \bibinfo {pages} {33} (\bibinfo {year}
  {2011})}\BibitemShut {NoStop}%
\bibitem [{\citenamefont {D'Hondt}\ and\ \citenamefont
  {Panangaden}(2004)}]{d2004computational}%
  \BibitemOpen
  \bibfield  {author} {\bibinfo {author} {\bibfnamefont {E.}~\bibnamefont
  {D'Hondt}}\ and\ \bibinfo {author} {\bibfnamefont {P.}~\bibnamefont
  {Panangaden}},\ }\href@noop {} {\bibfield  {journal} {\bibinfo  {journal}
  {arXiv preprint quant-ph/0412177}\ } (\bibinfo {year} {2004})}\BibitemShut
  {NoStop}%
\bibitem [{\citenamefont {Preskill}(2018)}]{Preskill2018quantumcomputingin}%
  \BibitemOpen
  \bibfield  {author} {\bibinfo {author} {\bibfnamefont {J.}~\bibnamefont
  {Preskill}},\ }\href {\doibase 10.22331/q-2018-08-06-79} {\bibfield
  {journal} {\bibinfo  {journal} {{Quantum}}\ }\textbf {\bibinfo {volume}
  {2}},\ \bibinfo {pages} {79} (\bibinfo {year} {2018})}\BibitemShut {NoStop}%
\bibitem [{\citenamefont {Galiautdinov}(2012)}]{galiautdinov2012simple}%
  \BibitemOpen
  \bibfield  {author} {\bibinfo {author} {\bibfnamefont {A.}~\bibnamefont
  {Galiautdinov}},\ }\href@noop {} {\bibfield  {journal} {\bibinfo  {journal}
  {arXiv preprint arXiv:1203.5534}\ } (\bibinfo {year} {2012})}\BibitemShut
  {NoStop}%
\bibitem [{\citenamefont {Kang}\ \emph {et~al.}(2016)\citenamefont {Kang},
  \citenamefont {Chen}, \citenamefont {Wu}, \citenamefont {Huang},
  \citenamefont {Song},\ and\ \citenamefont {Xia}}]{kang2016fast}%
  \BibitemOpen
  \bibfield  {author} {\bibinfo {author} {\bibfnamefont {Y.-H.}\ \bibnamefont
  {Kang}}, \bibinfo {author} {\bibfnamefont {Y.-H.}\ \bibnamefont {Chen}},
  \bibinfo {author} {\bibfnamefont {Q.-C.}\ \bibnamefont {Wu}}, \bibinfo
  {author} {\bibfnamefont {B.-H.}\ \bibnamefont {Huang}}, \bibinfo {author}
  {\bibfnamefont {J.}~\bibnamefont {Song}}, \ and\ \bibinfo {author}
  {\bibfnamefont {Y.}~\bibnamefont {Xia}},\ }\href@noop {} {\bibfield
  {journal} {\bibinfo  {journal} {Scientific reports}\ }\textbf {\bibinfo
  {volume} {6}},\ \bibinfo {pages} {36737} (\bibinfo {year}
  {2016})}\BibitemShut {NoStop}%
\bibitem [{\citenamefont {Xiao-Fang}\ and\ \citenamefont
  {Mei-Feng}(2011)}]{xiao2011generating}%
  \BibitemOpen
  \bibfield  {author} {\bibinfo {author} {\bibfnamefont {Y.}~\bibnamefont
  {Xiao-Fang}}\ and\ \bibinfo {author} {\bibfnamefont {C.}~\bibnamefont
  {Mei-Feng}},\ }\href@noop {} {\bibfield  {journal} {\bibinfo  {journal}
  {Communications in Theoretical Physics}\ }\textbf {\bibinfo {volume} {55}},\
  \bibinfo {pages} {868} (\bibinfo {year} {2011})}\BibitemShut {NoStop}%
\bibitem [{\citenamefont {Neeley}\ \emph {et~al.}(2010)\citenamefont {Neeley},
  \citenamefont {Bialczak}, \citenamefont {Lenander}, \citenamefont {Lucero},
  \citenamefont {Mariantoni}, \citenamefont {O’connell}, \citenamefont
  {Sank}, \citenamefont {Wang}, \citenamefont {Weides}, \citenamefont {Wenner}
  \emph {et~al.}}]{neeley2010generation}%
  \BibitemOpen
  \bibfield  {author} {\bibinfo {author} {\bibfnamefont {M.}~\bibnamefont
  {Neeley}}, \bibinfo {author} {\bibfnamefont {R.~C.}\ \bibnamefont
  {Bialczak}}, \bibinfo {author} {\bibfnamefont {M.}~\bibnamefont {Lenander}},
  \bibinfo {author} {\bibfnamefont {E.}~\bibnamefont {Lucero}}, \bibinfo
  {author} {\bibfnamefont {M.}~\bibnamefont {Mariantoni}}, \bibinfo {author}
  {\bibfnamefont {A.}~\bibnamefont {O’connell}}, \bibinfo {author}
  {\bibfnamefont {D.}~\bibnamefont {Sank}}, \bibinfo {author} {\bibfnamefont
  {H.}~\bibnamefont {Wang}}, \bibinfo {author} {\bibfnamefont {M.}~\bibnamefont
  {Weides}}, \bibinfo {author} {\bibfnamefont {J.}~\bibnamefont {Wenner}},
  \emph {et~al.},\ }\href@noop {} {\bibfield  {journal} {\bibinfo  {journal}
  {Nature}\ }\textbf {\bibinfo {volume} {467}},\ \bibinfo {pages} {570}
  (\bibinfo {year} {2010})}\BibitemShut {NoStop}%
\bibitem [{\citenamefont {Kiesel}\ \emph {et~al.}(2003)\citenamefont {Kiesel},
  \citenamefont {Bourennane}, \citenamefont {Kurtsiefer}, \citenamefont
  {Weinfurter}, \citenamefont {Kaszlikowski}, \citenamefont {Laskowski},\ and\
  \citenamefont {Zukowski}}]{kiesel2003three}%
  \BibitemOpen
  \bibfield  {author} {\bibinfo {author} {\bibfnamefont {N.}~\bibnamefont
  {Kiesel}}, \bibinfo {author} {\bibfnamefont {M.}~\bibnamefont {Bourennane}},
  \bibinfo {author} {\bibfnamefont {C.}~\bibnamefont {Kurtsiefer}}, \bibinfo
  {author} {\bibfnamefont {H.}~\bibnamefont {Weinfurter}}, \bibinfo {author}
  {\bibfnamefont {D.}~\bibnamefont {Kaszlikowski}}, \bibinfo {author}
  {\bibfnamefont {W.}~\bibnamefont {Laskowski}}, \ and\ \bibinfo {author}
  {\bibfnamefont {M.}~\bibnamefont {Zukowski}},\ }\href@noop {} {\bibfield
  {journal} {\bibinfo  {journal} {Journal of Modern Optics}\ }\textbf {\bibinfo
  {volume} {50}},\ \bibinfo {pages} {1131} (\bibinfo {year}
  {2003})}\BibitemShut {NoStop}%
\bibitem [{\citenamefont {Shi}\ and\ \citenamefont
  {Tomita}(2002)}]{shi2002schemes}%
  \BibitemOpen
  \bibfield  {author} {\bibinfo {author} {\bibfnamefont {B.-S.}\ \bibnamefont
  {Shi}}\ and\ \bibinfo {author} {\bibfnamefont {A.}~\bibnamefont {Tomita}},\
  }\href@noop {} {\bibfield  {journal} {\bibinfo  {journal} {arXiv preprint
  quant-ph/0208170}\ } (\bibinfo {year} {2002})}\BibitemShut {NoStop}%
\bibitem [{\citenamefont {Zou}\ \emph {et~al.}(2002)\citenamefont {Zou},
  \citenamefont {Pahlke},\ and\ \citenamefont {Mathis}}]{zou2002generation}%
  \BibitemOpen
  \bibfield  {author} {\bibinfo {author} {\bibfnamefont {X.}~\bibnamefont
  {Zou}}, \bibinfo {author} {\bibfnamefont {K.}~\bibnamefont {Pahlke}}, \ and\
  \bibinfo {author} {\bibfnamefont {W.}~\bibnamefont {Mathis}},\ }\href@noop {}
  {\bibfield  {journal} {\bibinfo  {journal} {Physical Review A}\ }\textbf
  {\bibinfo {volume} {66}},\ \bibinfo {pages} {044302} (\bibinfo {year}
  {2002})}\BibitemShut {NoStop}%
\bibitem [{\citenamefont {Li}\ and\ \citenamefont
  {Kobayashi}(2004)}]{li2004four}%
  \BibitemOpen
  \bibfield  {author} {\bibinfo {author} {\bibfnamefont {Y.}~\bibnamefont
  {Li}}\ and\ \bibinfo {author} {\bibfnamefont {T.}~\bibnamefont {Kobayashi}},\
  }\href@noop {} {\bibfield  {journal} {\bibinfo  {journal} {Physical Review
  A}\ }\textbf {\bibinfo {volume} {70}},\ \bibinfo {pages} {014301} (\bibinfo
  {year} {2004})}\BibitemShut {NoStop}%
\bibitem [{\citenamefont {Tashima}\ \emph {et~al.}(2008)\citenamefont
  {Tashima}, \citenamefont {{\"O}zdemir}, \citenamefont {Yamamoto},
  \citenamefont {Koashi},\ and\ \citenamefont {Imoto}}]{tashima2008elementary}%
  \BibitemOpen
  \bibfield  {author} {\bibinfo {author} {\bibfnamefont {T.}~\bibnamefont
  {Tashima}}, \bibinfo {author} {\bibfnamefont {{\c{S}}.~K.}\ \bibnamefont
  {{\"O}zdemir}}, \bibinfo {author} {\bibfnamefont {T.}~\bibnamefont
  {Yamamoto}}, \bibinfo {author} {\bibfnamefont {M.}~\bibnamefont {Koashi}}, \
  and\ \bibinfo {author} {\bibfnamefont {N.}~\bibnamefont {Imoto}},\
  }\href@noop {} {\bibfield  {journal} {\bibinfo  {journal} {Physical Review
  A}\ }\textbf {\bibinfo {volume} {77}},\ \bibinfo {pages} {030302} (\bibinfo
  {year} {2008})}\BibitemShut {NoStop}%
\bibitem [{\citenamefont {Heo}\ \emph {et~al.}(2019)\citenamefont {Heo},
  \citenamefont {Hong}, \citenamefont {Choi},\ and\ \citenamefont
  {Hong}}]{heo2019scheme}%
  \BibitemOpen
  \bibfield  {author} {\bibinfo {author} {\bibfnamefont {J.}~\bibnamefont
  {Heo}}, \bibinfo {author} {\bibfnamefont {C.}~\bibnamefont {Hong}}, \bibinfo
  {author} {\bibfnamefont {S.-G.}\ \bibnamefont {Choi}}, \ and\ \bibinfo
  {author} {\bibfnamefont {J.-P.}\ \bibnamefont {Hong}},\ }\href@noop {}
  {\bibfield  {journal} {\bibinfo  {journal} {Scientific reports}\ }\textbf
  {\bibinfo {volume} {9}},\ \bibinfo {pages} {1} (\bibinfo {year}
  {2019})}\BibitemShut {NoStop}%
\bibitem [{\citenamefont {H{\"a}ffner}\ \emph {et~al.}(2005)\citenamefont
  {H{\"a}ffner}, \citenamefont {H{\"a}nsel}, \citenamefont {Roos},
  \citenamefont {Benhelm}, \citenamefont {Chwalla}, \citenamefont {K{\"o}rber},
  \citenamefont {Rapol}, \citenamefont {Riebe}, \citenamefont {Schmidt},
  \citenamefont {Becher} \emph {et~al.}}]{haffner2005scalable}%
  \BibitemOpen
  \bibfield  {author} {\bibinfo {author} {\bibfnamefont {H.}~\bibnamefont
  {H{\"a}ffner}}, \bibinfo {author} {\bibfnamefont {W.}~\bibnamefont
  {H{\"a}nsel}}, \bibinfo {author} {\bibfnamefont {C.}~\bibnamefont {Roos}},
  \bibinfo {author} {\bibfnamefont {J.}~\bibnamefont {Benhelm}}, \bibinfo
  {author} {\bibfnamefont {M.}~\bibnamefont {Chwalla}}, \bibinfo {author}
  {\bibfnamefont {T.}~\bibnamefont {K{\"o}rber}}, \bibinfo {author}
  {\bibfnamefont {U.}~\bibnamefont {Rapol}}, \bibinfo {author} {\bibfnamefont
  {M.}~\bibnamefont {Riebe}}, \bibinfo {author} {\bibfnamefont
  {P.}~\bibnamefont {Schmidt}}, \bibinfo {author} {\bibfnamefont
  {C.}~\bibnamefont {Becher}},  \emph {et~al.},\ }\href@noop {} {\bibfield
  {journal} {\bibinfo  {journal} {Nature}\ }\textbf {\bibinfo {volume} {438}},\
  \bibinfo {pages} {643} (\bibinfo {year} {2005})}\BibitemShut {NoStop}%
\bibitem [{\citenamefont {Sharma}\ \emph {et~al.}(2008)\citenamefont {Sharma},
  \citenamefont {de~Almeida},\ and\ \citenamefont {Sharma}}]{Sharma_2008}%
  \BibitemOpen
  \bibfield  {author} {\bibinfo {author} {\bibfnamefont {S.~S.}\ \bibnamefont
  {Sharma}}, \bibinfo {author} {\bibfnamefont {E.}~\bibnamefont {de~Almeida}},
  \ and\ \bibinfo {author} {\bibfnamefont {N.~K.}\ \bibnamefont {Sharma}},\
  }\href {\doibase 10.1088/0953-4075/41/16/165503} {\bibfield  {journal}
  {\bibinfo  {journal} {Journal of Physics B: Atomic, Molecular and Optical
  Physics}\ }\textbf {\bibinfo {volume} {41}},\ \bibinfo {pages} {165503}
  (\bibinfo {year} {2008})}\BibitemShut {NoStop}%
\bibitem [{\citenamefont {Yu}\ and\ \citenamefont
  {Ying}(2013)}]{yu2013optimal}%
  \BibitemOpen
  \bibfield  {author} {\bibinfo {author} {\bibfnamefont {N.}~\bibnamefont
  {Yu}}\ and\ \bibinfo {author} {\bibfnamefont {M.}~\bibnamefont {Ying}},\
  }\href@noop {} {\bibfield  {journal} {\bibinfo  {journal} {arXiv preprint
  arXiv:1301.3727}\ } (\bibinfo {year} {2013})}\BibitemShut {NoStop}%
\bibitem [{\citenamefont {Diker}(2016)}]{diker2016deterministic}%
  \BibitemOpen
  \bibfield  {author} {\bibinfo {author} {\bibfnamefont {F.}~\bibnamefont
  {Diker}},\ }\href@noop {} {\bibfield  {journal} {\bibinfo  {journal} {arXiv
  preprint arXiv:1606.09290}\ } (\bibinfo {year} {2016})}\BibitemShut {NoStop}%
\bibitem [{\citenamefont {Martinez}\ \emph {et~al.}(2016)\citenamefont
  {Martinez}, \citenamefont {Monz}, \citenamefont {Nigg}, \citenamefont
  {Schindler},\ and\ \citenamefont {Blatt}}]{martinez2016compiling}%
  \BibitemOpen
  \bibfield  {author} {\bibinfo {author} {\bibfnamefont {E.~A.}\ \bibnamefont
  {Martinez}}, \bibinfo {author} {\bibfnamefont {T.}~\bibnamefont {Monz}},
  \bibinfo {author} {\bibfnamefont {D.}~\bibnamefont {Nigg}}, \bibinfo {author}
  {\bibfnamefont {P.}~\bibnamefont {Schindler}}, \ and\ \bibinfo {author}
  {\bibfnamefont {R.}~\bibnamefont {Blatt}},\ }\href@noop {} {\bibfield
  {journal} {\bibinfo  {journal} {New Journal of Physics}\ }\textbf {\bibinfo
  {volume} {18}},\ \bibinfo {pages} {063029} (\bibinfo {year}
  {2016})}\BibitemShut {NoStop}%
\bibitem [{\citenamefont {Bandyopadhyay}\ and\ \citenamefont
  {Cahay}(2015)}]{bandyopadhyay2015introduction}%
  \BibitemOpen
  \bibfield  {author} {\bibinfo {author} {\bibfnamefont {S.}~\bibnamefont
  {Bandyopadhyay}}\ and\ \bibinfo {author} {\bibfnamefont {M.}~\bibnamefont
  {Cahay}},\ }\href@noop {} {\emph {\bibinfo {title} {Introduction to
  spintronics}}}\ (\bibinfo  {publisher} {CRC press},\ \bibinfo {year}
  {2015})\BibitemShut {NoStop}%
\bibitem [{\citenamefont {Slonczewski}(1996)}]{slonczewski1996current}%
  \BibitemOpen
  \bibfield  {author} {\bibinfo {author} {\bibfnamefont {J.~C.}\ \bibnamefont
  {Slonczewski}},\ }\href@noop {} {\bibfield  {journal} {\bibinfo  {journal}
  {Journal of Magnetism and Magnetic Materials}\ }\textbf {\bibinfo {volume}
  {159}},\ \bibinfo {pages} {L1} (\bibinfo {year} {1996})}\BibitemShut
  {NoStop}%
\bibitem [{\citenamefont {Berger}(1996)}]{berger1996emission}%
  \BibitemOpen
  \bibfield  {author} {\bibinfo {author} {\bibfnamefont {L.}~\bibnamefont
  {Berger}},\ }\href@noop {} {\bibfield  {journal} {\bibinfo  {journal}
  {Physical Review B}\ }\textbf {\bibinfo {volume} {54}},\ \bibinfo {pages}
  {9353} (\bibinfo {year} {1996})}\BibitemShut {NoStop}%
\bibitem [{\citenamefont {Bhuktare}\ \emph {et~al.}(2019)\citenamefont
  {Bhuktare}, \citenamefont {Shukla}, \citenamefont {Singh}, \citenamefont
  {Bose},\ and\ \citenamefont {Tulapurkar}}]{bhuktare2019direct}%
  \BibitemOpen
  \bibfield  {author} {\bibinfo {author} {\bibfnamefont {S.}~\bibnamefont
  {Bhuktare}}, \bibinfo {author} {\bibfnamefont {A.~S.}\ \bibnamefont
  {Shukla}}, \bibinfo {author} {\bibfnamefont {H.}~\bibnamefont {Singh}},
  \bibinfo {author} {\bibfnamefont {A.}~\bibnamefont {Bose}}, \ and\ \bibinfo
  {author} {\bibfnamefont {A.~A.}\ \bibnamefont {Tulapurkar}},\ }\href@noop {}
  {\bibfield  {journal} {\bibinfo  {journal} {Applied Physics Letters}\
  }\textbf {\bibinfo {volume} {114}},\ \bibinfo {pages} {052402} (\bibinfo
  {year} {2019})}\BibitemShut {NoStop}%
\bibitem [{\citenamefont {Bose}\ \emph {et~al.}(2017)\citenamefont {Bose},
  \citenamefont {Dutta}, \citenamefont {Bhuktare}, \citenamefont {Singh},\ and\
  \citenamefont {Tulapurkar}}]{bose2017sensitive}%
  \BibitemOpen
  \bibfield  {author} {\bibinfo {author} {\bibfnamefont {A.}~\bibnamefont
  {Bose}}, \bibinfo {author} {\bibfnamefont {S.}~\bibnamefont {Dutta}},
  \bibinfo {author} {\bibfnamefont {S.}~\bibnamefont {Bhuktare}}, \bibinfo
  {author} {\bibfnamefont {H.}~\bibnamefont {Singh}}, \ and\ \bibinfo {author}
  {\bibfnamefont {A.~A.}\ \bibnamefont {Tulapurkar}},\ }\href@noop {}
  {\bibfield  {journal} {\bibinfo  {journal} {Applied Physics Letters}\
  }\textbf {\bibinfo {volume} {111}},\ \bibinfo {pages} {162405} (\bibinfo
  {year} {2017})}\BibitemShut {NoStop}%
\bibitem [{\citenamefont {Bose}\ \emph
  {et~al.}(2018{\natexlab{a}})\citenamefont {Bose}, \citenamefont {Lam},
  \citenamefont {Bhuktare}, \citenamefont {Dutta}, \citenamefont {Singh},
  \citenamefont {Jibiki}, \citenamefont {Goto}, \citenamefont {Miwa},\ and\
  \citenamefont {Tulapurkar}}]{bose2018observation}%
  \BibitemOpen
  \bibfield  {author} {\bibinfo {author} {\bibfnamefont {A.}~\bibnamefont
  {Bose}}, \bibinfo {author} {\bibfnamefont {D.~D.}\ \bibnamefont {Lam}},
  \bibinfo {author} {\bibfnamefont {S.}~\bibnamefont {Bhuktare}}, \bibinfo
  {author} {\bibfnamefont {S.}~\bibnamefont {Dutta}}, \bibinfo {author}
  {\bibfnamefont {H.}~\bibnamefont {Singh}}, \bibinfo {author} {\bibfnamefont
  {Y.}~\bibnamefont {Jibiki}}, \bibinfo {author} {\bibfnamefont
  {M.}~\bibnamefont {Goto}}, \bibinfo {author} {\bibfnamefont {S.}~\bibnamefont
  {Miwa}}, \ and\ \bibinfo {author} {\bibfnamefont {A.}~\bibnamefont
  {Tulapurkar}},\ }\href@noop {} {\bibfield  {journal} {\bibinfo  {journal}
  {Physical Review Applied}\ }\textbf {\bibinfo {volume} {9}},\ \bibinfo
  {pages} {064026} (\bibinfo {year} {2018}{\natexlab{a}})}\BibitemShut
  {NoStop}%
\bibitem [{\citenamefont {Bose}\ \emph {et~al.}(2016)\citenamefont {Bose},
  \citenamefont {Shukla}, \citenamefont {Konishi}, \citenamefont {Jain},
  \citenamefont {Asam}, \citenamefont {Bhuktare}, \citenamefont {Singh},
  \citenamefont {Lam}, \citenamefont {Fujii}, \citenamefont {Miwa} \emph
  {et~al.}}]{bose2016observation}%
  \BibitemOpen
  \bibfield  {author} {\bibinfo {author} {\bibfnamefont {A.}~\bibnamefont
  {Bose}}, \bibinfo {author} {\bibfnamefont {A.~K.}\ \bibnamefont {Shukla}},
  \bibinfo {author} {\bibfnamefont {K.}~\bibnamefont {Konishi}}, \bibinfo
  {author} {\bibfnamefont {S.}~\bibnamefont {Jain}}, \bibinfo {author}
  {\bibfnamefont {N.}~\bibnamefont {Asam}}, \bibinfo {author} {\bibfnamefont
  {S.}~\bibnamefont {Bhuktare}}, \bibinfo {author} {\bibfnamefont
  {H.}~\bibnamefont {Singh}}, \bibinfo {author} {\bibfnamefont {D.~D.}\
  \bibnamefont {Lam}}, \bibinfo {author} {\bibfnamefont {Y.}~\bibnamefont
  {Fujii}}, \bibinfo {author} {\bibfnamefont {S.}~\bibnamefont {Miwa}},  \emph
  {et~al.},\ }\href@noop {} {\bibfield  {journal} {\bibinfo  {journal} {Applied
  Physics Letters}\ }\textbf {\bibinfo {volume} {109}},\ \bibinfo {pages}
  {032406} (\bibinfo {year} {2016})}\BibitemShut {NoStop}%
\bibitem [{\citenamefont {Bose}\ and\ \citenamefont
  {Tulapurkar}(2019)}]{bose2019recent}%
  \BibitemOpen
  \bibfield  {author} {\bibinfo {author} {\bibfnamefont {A.}~\bibnamefont
  {Bose}}\ and\ \bibinfo {author} {\bibfnamefont {A.~A.}\ \bibnamefont
  {Tulapurkar}},\ }\href@noop {} {\bibfield  {journal} {\bibinfo  {journal}
  {Journal of Magnetism and Magnetic Materials}\ ,\ \bibinfo {pages} {165526}}
  (\bibinfo {year} {2019})}\BibitemShut {NoStop}%
\bibitem [{\citenamefont {Bose}\ \emph
  {et~al.}(2018{\natexlab{b}})\citenamefont {Bose}, \citenamefont {Bhuktare},
  \citenamefont {Singh}, \citenamefont {Dutta}, \citenamefont {Achanta},\ and\
  \citenamefont {Tulapurkar}}]{bose2018direct}%
  \BibitemOpen
  \bibfield  {author} {\bibinfo {author} {\bibfnamefont {A.}~\bibnamefont
  {Bose}}, \bibinfo {author} {\bibfnamefont {S.}~\bibnamefont {Bhuktare}},
  \bibinfo {author} {\bibfnamefont {H.}~\bibnamefont {Singh}}, \bibinfo
  {author} {\bibfnamefont {S.}~\bibnamefont {Dutta}}, \bibinfo {author}
  {\bibfnamefont {V.}~\bibnamefont {Achanta}}, \ and\ \bibinfo {author}
  {\bibfnamefont {A.}~\bibnamefont {Tulapurkar}},\ }\href@noop {} {\bibfield
  {journal} {\bibinfo  {journal} {Applied Physics Letters}\ }\textbf {\bibinfo
  {volume} {112}},\ \bibinfo {pages} {162401} (\bibinfo {year}
  {2018}{\natexlab{b}})}\BibitemShut {NoStop}%
\bibitem [{\citenamefont {Cordourier-Maruri}\ \emph {et~al.}(2010)\citenamefont
  {Cordourier-Maruri}, \citenamefont {Ciccarello}, \citenamefont {Omar},
  \citenamefont {Zarcone}, \citenamefont {De~Coss},\ and\ \citenamefont
  {Bose}}]{cordourier2010implementing}%
  \BibitemOpen
  \bibfield  {author} {\bibinfo {author} {\bibfnamefont {G.}~\bibnamefont
  {Cordourier-Maruri}}, \bibinfo {author} {\bibfnamefont {F.}~\bibnamefont
  {Ciccarello}}, \bibinfo {author} {\bibfnamefont {Y.}~\bibnamefont {Omar}},
  \bibinfo {author} {\bibfnamefont {M.}~\bibnamefont {Zarcone}}, \bibinfo
  {author} {\bibfnamefont {R.}~\bibnamefont {De~Coss}}, \ and\ \bibinfo
  {author} {\bibfnamefont {S.}~\bibnamefont {Bose}},\ }\href@noop {} {\bibfield
   {journal} {\bibinfo  {journal} {Physical Review A}\ }\textbf {\bibinfo
  {volume} {82}},\ \bibinfo {pages} {052313} (\bibinfo {year}
  {2010})}\BibitemShut {NoStop}%
\bibitem [{\citenamefont {Sutton}\ and\ \citenamefont
  {Datta}(2015)}]{sutton2015manipulating}%
  \BibitemOpen
  \bibfield  {author} {\bibinfo {author} {\bibfnamefont {B.}~\bibnamefont
  {Sutton}}\ and\ \bibinfo {author} {\bibfnamefont {S.}~\bibnamefont {Datta}},\
  }\href@noop {} {\bibfield  {journal} {\bibinfo  {journal} {Scientific
  reports}\ }\textbf {\bibinfo {volume} {5}},\ \bibinfo {pages} {17912}
  (\bibinfo {year} {2015})}\BibitemShut {NoStop}%
\bibitem [{\citenamefont {Kulkarni}\ \emph {et~al.}(2018)\citenamefont
  {Kulkarni}, \citenamefont {Prajapati},\ and\ \citenamefont
  {Kaushik}}]{kulkarni2018transmission}%
  \BibitemOpen
  \bibfield  {author} {\bibinfo {author} {\bibfnamefont {A.}~\bibnamefont
  {Kulkarni}}, \bibinfo {author} {\bibfnamefont {S.}~\bibnamefont {Prajapati}},
  \ and\ \bibinfo {author} {\bibfnamefont {B.~K.}\ \bibnamefont {Kaushik}},\
  }\href@noop {} {\bibfield  {journal} {\bibinfo  {journal} {IEEE Transactions
  on Very Large Scale Integration (VLSI) Systems}\ }\textbf {\bibinfo {volume}
  {26}},\ \bibinfo {pages} {1461} (\bibinfo {year} {2018})}\BibitemShut
  {NoStop}%
\bibitem [{\citenamefont {Kulkarni}\ and\ \citenamefont
  {Kaushik}(2019)}]{kulkarni2019spin}%
  \BibitemOpen
  \bibfield  {author} {\bibinfo {author} {\bibfnamefont {A.}~\bibnamefont
  {Kulkarni}}\ and\ \bibinfo {author} {\bibfnamefont {B.~K.}\ \bibnamefont
  {Kaushik}},\ }\href@noop {} {\bibfield  {journal} {\bibinfo  {journal} {IEEE
  Transactions on Magnetics}\ }\textbf {\bibinfo {volume} {55}},\ \bibinfo
  {pages} {1} (\bibinfo {year} {2019})}\BibitemShut {NoStop}%
\bibitem [{\citenamefont {Parkin}\ and\ \citenamefont
  {Mauri}(1991)}]{parkin1991spin}%
  \BibitemOpen
  \bibfield  {author} {\bibinfo {author} {\bibfnamefont {S.}~\bibnamefont
  {Parkin}}\ and\ \bibinfo {author} {\bibfnamefont {D.}~\bibnamefont {Mauri}},\
  }\href@noop {} {\bibfield  {journal} {\bibinfo  {journal} {Physical Review
  B}\ }\textbf {\bibinfo {volume} {44}},\ \bibinfo {pages} {7131} (\bibinfo
  {year} {1991})}\BibitemShut {NoStop}%
\end{thebibliography}%

\end{document}